\documentclass[a4paper,11pt]{article}
\usepackage[utf8]{inputenc}
\usepackage[table]{xcolor}
\usepackage{jcappub}
\usepackage{graphicx}
\usepackage{psfrag,fancyhdr,epsfig}
\usepackage{hyperref}

\hypersetup{colorlinks,bookmarksopen,bookmarksnumbered,citecolor=magenta,linkcolor=blue,pdfstartview=FitH,urlcolor=blue}
\usepackage{cancel}
\usepackage{soul}
\usepackage{amsmath}
\usepackage{array}
\usepackage{subfig}
\usepackage{color}
\usepackage{tensor}
\usepackage{orcidlink}
\usepackage{comment}
\usepackage{physics}
\usepackage[mathscr]{eucal}
\usepackage{multirow}
\usepackage{amssymb}
\newcommand{\cd}{\mathring{\nabla}}
\newcommand{\lcd}{\mathring{\nabla}}          
\newcommand{\lcR}{\mathring{R}}               
\newcommand{\lcG}{\mathring{\Gamma}}          

\newcommand{\Lalg}{\mathcal{L}_{\rm alg}}

\newcommand{\Mp}{M_P}

\newcommand{\Qt}{\tilde{Q}}

\usepackage{comment}

\begin{document}
\title{\centering{Inflationary attractors in\\ metric-affine gravity}}
\author[a]{Ilaria Andrei,}
\author[b]{Christian Dioguardi \orcidlink{0000-0002-6133-0383}}
\affiliation[a]{Laboratory of Theoretical Physics, Institute of Physics, University of Tartu, W. Ostwaldi 1, 50411 Tartu, Estonia.}
\affiliation[b]{Laboratory of High Energy and Computational Physics,
National Institute of Chemical Physics and Biophysics, R{\"a}vala pst.~10, Tallinn, 10143, Estonia}
\emailAdd{ilaria.andrei@ut.ee}
\emailAdd{christian.dioguardi@kbfi.ee}

\abstract{We study single-field slow-roll inflation in metric-affine gravity (MAG). We consider the most general action up to operators linear in curvature and quadratic in the distortion, without propagating torsion and non-metricity. 
We first impose projective invariance and promote every coefficient to a function of the scalar field. After integrating out the non-dynamical distortion, the net contribution is a correction $k(\phi)$ to the scalar kinetic term. We then consider the class of $\xi$-attractors for which the non-minimal couplings and the scalar potential are fixed by a single function $f(\phi)$. It is well-known that in the metric formulation the strong-coupling regime is universally the Starobinsky attractor. In the Palatini framework, instead, this attractor is lost and the predictions become model-dependent. Here, we show that the MAG framework restores universality: in the strong coupling limit with finite ratios of the distortion couplings to $\xi$ we have an $\alpha$-attractor, $n_s=1-2/N_*$ and $r=12\alpha/N_*^2$, with $\alpha=1+\zeta/(6\xi)$.
The ratio $\zeta/\xi$ encodes the distortion contribution.
Starobinsky and Palatini inflation are recovered as specific limits. 
Through a numerical parameter scan we show that almost all the healthy points lie on one of the two attractors 
: either Nieh-Yan-like inflation (studied in a previous work and extended here) or the new MAG $\alpha$-attractor found here. Finally,
we explicitly drop projective invariance and analyse the general MAG theory. The same large-field asymptotic structure survives. We conclude that no extra inflationary phenomenology is reached by the non-projectively invariant operators: the $\alpha$-attractor is a general prediction of MAG.}
\maketitle

\section{Introduction}

Homogeneity and flatness of the Universe are among the most precisely measured properties in cosmology
\cite{Planck:2018vyg}. However, within $\Lambda$CDM, which currently best describes the evolution of the Universe, these features enter as initial conditions rather than
predictions. An early epoch of quasi-de Sitter expansion, inflation, remains the most economical way of giving a dynamical explanation of homogeneity and flatness
\cite{Starobinsky:1980te, Guth:1980zm, Linde:1981mu, Albrecht:1982wi}. Moreover inflation provides, at no additional cost, the nearly scale-invariant spectrum of curvature
perturbations from which the large-scale structure later develops.\\
In the standard picture a single scalar degree of freedom, the inflaton, drives the
expansion, and the shape of its potential fixes, in the slow-roll approximation, the observables that the CMB constrains: the amplitude of the scalar perturbations $A_s$, the scalar spectral
index $n_s$ and the tensor-to-scalar ratio $r$ \cite{Martin:2013tda}. The observational
status of inflation has changed significantly in the last few years. The joint analysis of Planck,
BICEP/Keck, ACT and SPT data constrains $r\lesssim0.035$, at $95\%$ CL, $n_s = 0.9682 \pm 0.0032$ at $68\%$ CL
\cite{Planck:2018jri, BICEP:2021xfz, AtacamaCosmologyTelescope:2025nti, SPT-3G:2025bzu, Balkenhol:2025wms},
which by itself disfavours simple monomial potentials in a minimally coupled setting. At the same
time, adding the DESI BAO measurements to the combined CMB data moves the inferred spectral index
upwards, to $n_s=0.9728\pm0.0029$ at $68\%$ CL \cite{DESI:2024mwx, DESI:2025zgx,
AtacamaCosmologyTelescope:2025nti, SPT-3G:2025bzu, Balkenhol:2025wms}. This potentially leaves popular models such as Starobinsky \cite{Starobinsky:1980te} and Higgs \cite{Bezrukov:2007ep} inflation
$\gtrsim 2\sigma$ away from the preferred value of $n_s$,
while the $r\sim3\times10^{-3}$ they predict sits within the reach of the
next decade of CMB experiments \cite{SimonsObservatory:2018koc, LiteBIRD:2020khw,
SPT:2026olz}\footnote{The pull towards larger $n_s$ is largely driven by a $2$-$3\sigma$
tension between the CMB and BAO datasets \cite{Ferreira:2025lrd, Balkenhol:2025wms},
so it cannot yet be attributed unambiguously to the primordial spectrum. Accordingly, we
display throughout the paper the constraints obtained with and without DESI data.}.\\
Of the available mechanisms that suppress the tensor-to-scalar ratio $r$, a non-minimal coupling between the inflaton and the spacetime curvature
is one of the most economical. This class of models, however, is sensitive to the geometric assumptions about gravity. In the metric formulation the connection is not an independent
object: it is fixed to be the Levi-Civita connection of $g_{\mu\nu}$. If
instead the (symmetric) connection is varied independently, as in the Palatini approach, the two
descriptions cease to be equivalent the moment the scalar couples non-minimally, and the resulting
predictions can differ dramatically \cite{Bauer:2008zj, Marzola:2016xgb, Jarv:2017azx,
Racioppi:2018zoy, Enckell:2018hmo, Jarv:2020qqm, Gialamas:2020snr, Dioguardi:2021fmr,
Kallosh:2022feu, Dioguardi:2022oqu, Dioguardi:2023jwa, Jarv:2024krk,Bostan:2025zdt, Dioguardi:2025mpp,Dioguardi:2025vci}.\\

Metric-affine gravity (MAG) \cite{Hehl:1976my, Hehl:1976kj, Hehl:1994ue} is the natural
extension of this line of reasoning: the connection is left entirely free, neither metric-compatible nor symmetric, so that its departure from Levi-Civita is measured by two additional geometric properties of spacetime: non-metricity $Q_{\rho\mu\nu}$ and torsion $T^\rho{}_{\mu\nu}$. Varying the matter action with
respect to the independent connection defines the hypermomentum \cite{Hehl:1976kt, Hehl:1978cb}, which, together with the energy-momentum tensor, obtained from the variation with respect to the metric, has to be regarded as the source of the field equations.\\
Whether this setting is useful for inflation depends on what one does with the 
torsion and non-metricity components.
Letting them propagate produces a considerably richer spectrum \cite{Baldazzi:2021kaf}, but theories of this type are notoriously hard to keep
free of ghosts \cite{BeltranJimenez:2019acz, Percacci:2020ddy, Mikura:2023ruz,
Mikura:2024mji, Barker:2025xzd, Percacci:2025oxw}. A more restrictive possibility is that only a single scalar mode of the geometry becomes dynamical. This is what happens when the gravitational action goes beyond linear order in the curvature, e.g.\ with quadratic (or higher) powers of the Ricci and Holst scalars, or with a nonlinear function of the Ricci scalar supplemented by quadratic torsion and non-metricity invariants: the distortion can still be solved for, but integrating it out generates a kinetic term for the auxiliary scalar of the higher-curvature sector, which then propagates as an additional scalar degree of freedom. This scalar of gravitational origin can drive inflation on its own or together with a fundamental inflaton \cite{Salvio:2022suk, Gialamas:2022xtt, DiMarco:2023ncs, Gialamas:2024iyu, He:2024wqv, He:2025bli, Karananas:2025xcv, Salvio:2025izr,Iosifidis:2025wrv, Katsoulas:2025srh}. A notable, testable example has been recently studied in \cite{Dioguardi:2026hfe}, where the so-called $\mathring{R}^2$ scalaron is constructed from a symmetry principle and drives parameter-free inflation in very good agreement with current observations. See \cite{Capozziello:2024lsz} for a comparison of geometry-driven inflation across extended metric, teleparallel and symmetric-teleparallel theories.\\
In this work we take the opposite route and keep the particle content minimal: the connection is non-dynamical, so that $T^{\rho}{}_{\mu\nu}$ and $Q_{\rho\mu\nu}$ enter the action only algebraically and on a non-degenerate branch, can be eliminated exactly. The only scalar degree of freedom is the inflaton itself. This is not the same as discarding the distortion: whenever the hypermomentum is non-zero the
connection equation is not solved by Levi-Civita, and substituting the solution back leaves
an imprint on the scalar sector, typically in the form of a modified kinetic term for the inflaton. After a conformal transformation of the metric and a canonical normalization of the scalar field this yields a reshaped Einstein-frame potential.\\
This mechanism, with a fundamental inflaton coupled to non-dynamical torsion and/or non-metricity, has been exploited extensively in several sub-cases \cite{Shimada:2018lnm, Shaposhnikov:2020gts,
Shaposhnikov:2020frq, Langvik:2020nrs, Mikura:2020qhc, Mikura:2021ldx, Piani:2022gon, Rigouzzo:2022yan,
Racioppi:2024zva, Gialamas:2024jeb, Racioppi:2024pno, Gialamas:2024uar, 
Racioppi:2025igu, Dimopoulos:2026iwq, Kraiko:2026nas, Hassan:2026ddx, Andrei:2026gmu}, the Holst and Nieh--Yan couplings being its most thoroughly studied instances. Closely related constructions have also
been applied to late-time cosmology \cite{Iosifidis:2020gth, Iosifidis:2021nra,
Andrei:2024vvy, Dimopoulos:2026iwq, Andrei:2026ryg}. 
Here, we consider the most general Lagrangian, parity-even and
parity-odd, that is linear in the curvature and quadratic in the distortion \cite{Hehl:1994ue, Iosifidis:2021bad}. Promoting every
coefficient to a function of the inflaton generates twenty-three arbitrary functions of the scalar field \cite{Andrei:2026xad}.
Imposing projective symmetry reduces them to eighteen.

In order to simplify computations and make results more readable, we first study this symmetry-defined sector. We then return to the full metric-affine theory in section~\ref{sec:MAG} and show that the same large-field attractor structure persists after projective invariance is dropped.

We find the following: since the only source available is the gradient of the scalar field $\partial_\mu\phi$, the distortion is forced into a
four-parameter family, of which one direction is pure projective gauge; the general solution
is therefore governed by three invariants, and the connection field equation collapses to a
$3\times3$ linear system. We verify, as a consistency check dictated by the Noether identity
associated with projective symmetry, that the dilation current of the hypermomentum vanishes
identically. Integrating the distortion out yields the induced correction $k(\phi)$ to the scalar kinetic term. \\

We then focus on the well-studied class of $\xi$-attractors, for which the non-minimal coupling to the Ricci scalar is $F(\phi)=M_P^2\qty(1+\xi f(\phi))$ and the potential is $V(\phi)=M^4f(\phi)^2$, with $M$ a constant mass scale. In MAG we extend this ansatz by taking all distortion couplings proportional to $f(\phi)$. It is well-known that in the metric formulation of gravity this class universally predicts the Starobinsky inflationary attractor \cite{Kallosh:2013maa}, in the strong coupling regime. It is also well-known that in the Palatini formulation the attractor is lost and the predictions become model dependent \cite{Dioguardi:2026dfw}.
We show here that, provided the kinetic function stays positive throughout field space, universality is restored at strong coupling regime with finite ratios of the distortion couplings to $\xi$: the predictions are those of an $\alpha$-attractor \cite{Kallosh:2013hoa,Ferrara:2013rsa,Kallosh:2013yoa,Kallosh:2013tua,Galante:2014ifa,Carrasco:2015uma} with
\begin{equation*}
n_s=1-\frac{2}{N_*},\qquad r=\frac{12\alpha}{N_*^2},\qquad \alpha=1+\frac{\zeta}{6\xi},
\end{equation*}
up to an $f(\phi)$-dependent correction, which vanishes when $f(\phi)$ grows faster than $\phi^2$ and equals $1/(6\xi)$ for $f(\phi)=\phi^2/M_P^2$. Here $N_*$ is the number of $e$-folds between horizon exit of the pivot scale and the end of inflation, and the single constant $\zeta$ carries the entire contribution of the metric-affine couplings. Starobinsky inflation is recovered for $\zeta/\xi\to0$, while Palatini inflation sits at $\zeta=-6\xi$, where only the $f(\phi)$-dependent term survives and universality is lost. Note that at leading order in $1/N_*$ the spectral index coincides with the Starobinsky one: when the attractor is reached the metric-affine couplings do not shift $n_s$, but turn $r$ into a free parameter at fixed $n_s$.
Positivity of the kinetic function is not guaranteed by projective invariance alone: a scan
over the fourteen dimensionless couplings shows that roughly only $25\%$ of the sampled points is ghost-free, while essentially every healthy point produces successful inflation and is a result of the competition between two attractors: one generated by Nieh--Yan-like operators and recently studied in \cite{Andrei:2026gmu}, and the metric-affine $\alpha$-attractor identified here. The latter contains the Starobinsky case at $\alpha_0=1$, whereas the Palatini limit lies on the hypersurface $\alpha_0=0$, where universality is lost.

Finally, we remove the assumption of projective invariance and consider the general metric-affine action. The scalar-sourced distortion then contains four independent modes and the connection equation enlarges from a $3\times3$ to a $4\times4$ algebraic system. Nevertheless, the asymptotic structure responsible for the attractor is unchanged: at large field values the distortion-induced kinetic correction 
leads to an $\alpha$-attractor with predictions independent of the specific choice of $f(\phi)$. This shows that the metric-affine $\alpha$-attractor is not a consequence of projective symmetry, but persists in the general MAG theory considered here. The additional non-projectively invariant operators mainly reduce the healthy region of parameter space, rather than generating a qualitatively new inflationary regime.

The paper is organised as follows. In section \ref{sec:setup} we set up the metric-affine
framework and fix conventions. Section \ref{sec:proj_inv} motivates projective invariance as
a selection principle and constructs the invariant building blocks out of torsion and
non-metricity. In section \ref{sec:action} we write down the most general projectively
invariant action that is linear in the curvature and quadratic in the distortion. Section \ref{sec:integrate}
solves the connection field equation, establishes the vanishing of the dilation current and
derives the induced kinetic function $k(\phi)$. Section \ref{sec:inflation}
passes to the Einstein-frame, reviews the slow-roll formalism, and demonstrates the emergence of
the $\alpha$-attractor in the large-field regime. In section \ref{sec:parameters} we compare the predictions with the current CMB data. We perform a scan of the fourteen dimensional parameter space and highlight the appearance of the two main classes of inflationary attractors that characterize inflation in MAG. In section \ref{sec:MAG} we drop projective invariance and show that no new phenomenology is reached by the non-projectively invariant operators: the $\alpha$-attractor is a general prediction of MAG. Finally, we summarize and conclude in section \ref{sec:conclusions}. Appendices \ref{app:coeff},\ref{app:coeff_2}
collects the explicit expressions for all the effective couplings and  coefficients used throughout this paper. Appendix \ref{app:coeff_3} contains a dictionary for comparison with a previous work \cite{Andrei:2026xad}.

\section{Metric-affine framework and conventions}\label{sec:setup}

We work on a four-dimensional non-Riemannian manifold $(\mathcal M,g_{\mu\nu},\Gamma^\rho{}_{\mu\nu})$
with signature $(-,+,+,+)$ and an independent affine connection. We decompose the affine connection as

\begin{equation}\label{eq:affine}
\Gamma^\rho{}_{\mu\nu}=\lcG^\rho{}_{\mu\nu}+N^\rho{}_{\mu\nu},
\end{equation}
where $\lcG$ is the Levi-Civita connection of $g_{\mu\nu}$ given by

\begin{equation}
    \label{eq:lc_conn}
    \lcG^{\rho}{}_{\mu\nu} = \frac{1}{2} g^{\rho\lambda}
    \left(\partial_\mu g_{\nu\lambda} + \partial_\nu g_{\lambda\mu} - \partial_\lambda g_{\mu\nu}\right),
\end{equation}
and $N^\rho{}_{\mu\nu}$ is the distortion tensor.
We use the convention in which the \emph{first} lower index of $\Gamma$ is the differentiation index:

\begin{equation}\label{eq:cov_d}
\nabla_\mu V^\rho=\partial_\mu V^\rho+\Gamma^\rho{}_{\mu\nu}V^\nu,
\qquad
\nabla_\mu \omega_\rho=\partial_\mu\omega_\rho-\Gamma^\lambda{}_{\mu\rho}\omega_\lambda .
\end{equation}
The distortion can be expressed in terms of two quantities that characterize the metric-affine geometry, torsion $T$ and non-metricity $Q$ defined as 
\begin{equation}\label{eq:TQdef}
T^\rho{}_{\mu\nu}:=2\Gamma^\rho{}_{[\mu\nu]},\qquad
  Q_{\rho\mu\nu}:=\nabla_\rho g_{\mu\nu},
\end{equation}
where $T_{\rho\mu\nu}=-T_{\rho\nu\mu}$ and $Q_{\rho\mu\nu}=Q_{\rho\nu\mu}$ are respectively antisymmetric and symmetric in the last pair of indices.
Indices are raised and lowered with $g_{\mu\nu}$. For torsion and distortion, we lower the originally contravariant first index according to $T_{\rho\mu\nu}\equiv g_{\rho\lambda}T^\lambda{}_{\mu\nu}$, $N_{\rho\mu\nu}\equiv g_{\rho\lambda}N^\lambda{}_{\mu\nu}$.
Thus, 
\begin{equation}
N_{\rho\mu\nu}=\tfrac12\big(Q_{\rho\mu\nu}-Q_{\nu\rho\mu}-Q_{\mu\rho\nu}\big)
                +\tfrac12\big(T_{\rho\mu\nu}+T_{\nu\rho\mu}+T_{\mu\rho\nu}\big).
  \label{eq:distortion}
\end{equation}
From torsion and non-metricity we can further define three independent traces: 
\begin{equation}\label{eq:traces}
  T_\mu=T^\nu{}_{\nu\mu},\qquad
  Q_\mu=Q_\mu{}^\nu{}_\nu,\qquad
  \tilde Q_\mu=Q^\nu{}_{\nu\mu}.
\end{equation}
The curvature of the affine connection is defined through the commutator of covariant derivatives acting on a vector $V^\mu$:
\begin{equation}\label{eq:commutator}
  [\nabla_\alpha,\nabla_\beta]V^\mu=R^\mu{}_{\nu\alpha\beta}V^\nu
  -T^\lambda{}_{\alpha\beta}\nabla_\lambda V^\mu ,
\end{equation}
where the non-Riemannian curvature tensor is 
\begin{equation}\label{eq:Riemanntensor}
R^\mu{}_{\nu\alpha\beta} := \partial_\alpha \Gamma^\mu{}_{\beta\nu}
- \partial_\beta \Gamma^\mu{}_{\alpha\nu}
+ \Gamma^\mu{}_{\alpha\lambda}\Gamma^\lambda{}_{\beta\nu}
- \Gamma^\mu{}_{\beta\lambda}\Gamma^\lambda{}_{\alpha\nu},
\end{equation}
constructed from the independent connection $\Gamma$. 

Unlike the Riemann tensor of $g_{\mu\nu}$, the tensor \eqref{eq:Riemanntensor} is antisymmetric only in its last two indices. Nevertheless, only two independent scalars linear in the curvature can be built from contractions with $g_{\mu\nu}$ and with the Levi-Civita tensor $\epsilon^{\mu\nu\rho\sigma}$ \footnote{Here $\epsilon^{\mu\nu\rho\sigma}\equiv-\tilde\epsilon^{\mu\nu\rho\sigma}/\sqrt{-g}$, with $\tilde\epsilon^{\mu\nu\rho\sigma}$ the totally antisymmetric symbol, $\tilde\epsilon^{0123}=1$.}. In fact, every full metric contraction reduces to one of three cases: $g^{\alpha\beta}R^\mu{}_{\mu\alpha\beta}=0$, since the homothetic curvature $R^\mu{}_{\mu\alpha\beta}$ is antisymmetric in $\alpha\beta$, while $g^{\alpha\beta}R^\mu{}_{\alpha\mu\beta}=R$ and $g^{\alpha\beta}R^\mu{}_{\alpha\beta\mu}=-R$ differ only by a sign. Adding the contraction with the Levi-Civita tensor as well, we have the two scalars
\begin{equation}\label{eq:twoscalars}
R:=g^{\mu\nu}R_{\mu\nu}=g^{\mu\nu}R^\sigma{}_{\mu\sigma\nu},\qquad
\tilde{R} := \epsilon^{\mu\nu\rho\sigma}R_{\mu\nu\rho\sigma},
\end{equation}
the Ricci scalar and the Holst invariant, respectively. Note that for the Levi-Civita connection $\tilde R$ vanishes identically by the first Bianchi identity.

Using \eqref{eq:affine}, \eqref{eq:distortion} and
\eqref{eq:Riemanntensor}, we can express the relation between the non-Riemannian Ricci scalar and the metric-dependent Ricci scalar:
\begin{equation}\label{eq:Rdecomp}
\begin{split}
R(\Gamma) = \lcR(g) 
&+\tfrac14\,T_{\rho\mu\nu}T^{\rho\mu\nu}
 +\tfrac12\,T_{\rho\mu\nu}T^{\mu\rho\nu}
+Q_{\rho\mu\nu}T^{\mu\rho\nu}+\tfrac14\,Q_{\rho\mu\nu}Q^{\rho\mu\nu}
 -\tfrac12\,Q_{\rho\mu\nu}Q^{\mu\rho\nu}\\
&-T_\mu T^\mu
 -\tfrac14\,Q_\mu Q^\mu
 +\tfrac12\,Q_\mu\tilde Q^\mu
 +Q_\mu T^\mu
 -\tilde Q_\mu T^\mu + \lcd_\mu X^\mu,
\end{split}
\end{equation}
with 
\begin{equation}
  X^\mu=Q^\mu-\tilde Q^\mu-2T^\mu,
  \label{eq:Xvec}
\end{equation}
and where $\lcd$ and $\lcR(g)$ denote the covariant derivative and the Ricci scalar of the Levi-Civita connection.
One can straightforwardly check that both $R(\Gamma)$ and $\tilde{R}(\Gamma)$ in \eqref{eq:twoscalars} are invariant under the projective transformation
\begin{equation}
  \Gamma^\rho{}_{\mu\nu}\ \rightarrow\ \Gamma^\rho{}_{\mu\nu}+\delta^\rho{}_\nu \theta_\mu,
  \label{eq:proj}
\end{equation}
with $\theta_\mu$ an arbitrary vector. 
In the following we will extensively discuss projective invariance with the aim of extending
this property to the full metric-affine action.

\section{Projective invariance}\label{sec:proj_inv}

The role of projective symmetry has been extensively discussed in the context of metric-affine gravity \cite{Afonso:2017bxr,BeltranJimenez:2019acz,BeltranJimenez:2020sqf,Klemm:2020mfp,Garcia-Parrado:2020lpt,Iosifidis:2018zwo,Iosifidis:2018zjj,Ferrara:2026bzy}. 
Projective symmetry can help eliminate ghost-like degrees of freedom associated with the projective mode and has been employed in constructing ghost-free classes of theories with dynamical torsion and non-metricity \cite{Aoki:2019rvi,BeltranJimenez:2020sqf,Percacci:2020ddy}.
However, this does not come for free. Imposing projective invariance leads to a Noether identity requiring the dilation current of the hypermomentum tensor to vanish (see discussion in section \ref{sec:linear_system}). A generic matter sector does not necessarily satisfy this condition, particularly in the presence of non-minimal couplings to the independent connection \footnote{Standard matter minimally coupled to the gravitational sector is actually projectively invariant as discussed in \cite{BeltranJimenez:2017doy,BeltranJimenez:2018vdo}, however this might not hold for non-minimally coupled fermions.}.
\\
Even in the case of non-dynamical torsion and non-metricity (which is the one considered in this paper) there are several good reasons to consider projective invariance.
The general metric-affine action, counting both parity-even and parity-odd terms, at most quadratic in the distortion contains seventeen
independent invariants~\cite{Hehl:1994ue,Iosifidis:2021bad}; once every coefficient is allowed
to depend on a non-minimally coupled scalar field, the theory contains twenty-three free functions (the 17 quadratic coefficients, the curvature coupling, the scalar potential and the linear derivative couplings to the four independent trace/axial vectors). Such a Lagrangian usually requires some organising principle before anything can be understood. The usual practice is to truncate by hand (keeping only the Holst and Nieh--Yan invariants, considering parity-even operators only, or setting either torsion or non-metricity to zero) which is economical but leaves open the question of what has been discarded. Imposing a symmetry instead selects a subspace of the coupling space for a stated reason, and one knows exactly which operators are removed and why.

Of all the operators available, only $R(\Gamma)$ and
$\tilde R(\Gamma)$ involve the connection through its curvature (at linear order) rather than algebraically, and
both are invariant under \eqref{eq:proj}, whereas a generic quadratic term in the distortion is not.
Imposing \eqref{eq:proj} on the whole action therefore extends to the algebraic sector a
property that the curvature sector possesses automatically.

Two features make this a useful selection rule. First, the resulting subspace contains as
special cases several of the most extensively studied models, in the context of non-minimally coupled scalar fields such as metric and Palatini Higgs inflation \cite{Bauer:2008zj,Rasanen:2017ivk,Racioppi:2019jsp}, inflation with the Holst and Nieh--Yan terms \cite{Langvik:2020nrs,Gialamas:2022xtt, Racioppi:2024zva} etc., so it is a symmetry-defined completion of the existing literature rather than an arbitrary enlargement of it. Second, it treats the projective vector coherently. In the projectively invariant curvature sector, the connection field equations leave the projective mode $\theta_\mu$ undetermined, while the metric field equations are insensitive to it ~\cite{Bernal:2016lhq}: two connections related by \eqref{eq:proj} describe the same physics,
so $\theta_\mu$ labels a redundancy of the description.

Requiring every operator in the action to be invariant under \eqref{eq:proj} ensures that connections related by a projective transformation are physically equivalent. We adopt it on those grounds, as a selection principle rather than as a consistency requirement.

Under \eqref{eq:proj} torsion and non-metricity transform as 
\begin{equation}
  T_{\rho\mu\nu}\rightarrow T_{\rho\mu\nu} + g_{\rho\nu}\theta_\mu-g_{\rho\mu}\theta_\nu,
  \qquad Q_{\rho\mu\nu}\rightarrow Q_{\rho\mu\nu}-2\,g_{\mu\nu}\theta_\rho,
  \label{eq:projTQ}
\end{equation}
so that 
\begin{equation}
T_\mu\rightarrow T_\mu-3 \theta_\mu,   \qquad
Q_\mu\rightarrow Q_\mu-8 \theta_\mu,   \qquad
\tilde Q_\mu\rightarrow \tilde Q_\mu-2 \theta_\mu.
\end{equation}
We now construct manifestly projectively invariant operators, working directly with $T_{\rho\mu\nu}$ and $Q_{\rho\mu\nu}$, which by \eqref{eq:distortion} carry the same information as $N_{\rho\mu\nu}$. Within each sector the shift \eqref{eq:projTQ} is pure trace: $\delta T_{\rho\mu\nu}=g_{\rho\nu}\theta_\mu-g_{\rho\mu}\theta_\nu$ has the tensor structure generated by $T_\mu$, and $\delta Q_{\rho\mu\nu}=-2g_{\mu\nu}\theta_\rho$ the one generated by $Q_\rho$. Each shift can therefore be compensated within its own sector by a single trace subtraction, with coefficients fixed by $\delta T_\mu=-3\theta_\mu$ and $\delta Q_\mu=-8\theta_\mu$, giving the invariant rank-three building blocks
\begin{equation}
  \mathcal T_{\rho\mu\nu}=T_{\rho\mu\nu}+\tfrac13\big(g_{\rho\nu}T_\mu-g_{\rho\mu}T_\nu\big),
  \qquad
  \mathcal Q_{\rho\mu\nu}=Q_{\rho\mu\nu}-\tfrac14 g_{\mu\nu}Q_\rho ,
  \label{eq:traceless}
\end{equation}
with $g^{\rho\nu}\mathcal T_{\rho\mu\nu}=0$ and $g^{\mu\nu}\mathcal Q_{\rho\mu\nu}=0$. The choice is not unique, since terms built from the invariant vectors introduced below can be added; \eqref{eq:traceless} is the one for which $\mathcal T_{\rho\mu\nu}$ is totally traceless. Mixed quadratic operators constructed from the projectively invariant tensors, such as $\mathcal Q^{\rho\mu\nu}\mathcal T_{\mu\rho\nu}$, are also admissible in the Lagrangian and will appear below. Note that while $\mathcal T_{\rho\mu\nu}$ is totally traceless, $\mathcal Q_{\rho\mu\nu}$ is traceless only on its last pair, its remaining trace being $g^{\rho\mu}\mathcal Q_{\rho\mu\nu}=U_\nu$ with $U_\nu$ defined below in Eq.~\eqref{eq:WU}.

Let us now consider rank one. The three trace
vectors \eqref{eq:traces} all shift proportionally to $\theta_\mu$, so they span a
three-dimensional space of which one direction is pure gauge, and exactly two parity-even
invariant combinations survive,
\begin{equation}
  W_\mu=T_\mu-\tfrac38 Q_\mu,\qquad U_\mu=\tilde Q_\mu-\tfrac14 Q_\mu ,
  \label{eq:WU}
\end{equation}
together with a single parity-odd invariant vector:
\begin{equation}
  \bar T^\mu:=\epsilon^{\mu\nu\rho\sigma}T_{\nu\rho\sigma}
        =\epsilon^{\mu\nu\rho\sigma}\mathcal T_{\nu\rho\sigma},
  \label{eq:axial}
\end{equation}
where the second equality holds because the trace part of $T_{\nu\rho\sigma}$ drops out of the
$\epsilon$ contraction. 

Since non-metricity is symmetric in its last two indices, it cannot generate an independent axial vector through contraction with the Levi-Civita tensor. Therefore, $\bar T^\mu$ \eqref{eq:axial} is the only independent parity-odd vector linear in torsion and non-metricity

Finally, the boundary vector appearing in \eqref{eq:Rdecomp} 
is projectively invariant but not independent:
\begin{equation}
  X^\mu=-\big(2W^\mu+U^\mu\big).
  \label{eq:XWU}
\end{equation}

\section{The projective-invariant action}\label{sec:action}

We retain all operators that are at most quadratic in the distortion and linear in the curvature, assuming a canonical kinetic term for the scalar field $\phi$.
The most general projectively invariant action non-minimally coupled to a real scalar then reads
\begin{align}
  \mathcal{S}=\int\!\dd^4x\sqrt{-g}\;\Big[\;
  &\tfrac12 F(\phi)\,R(\Gamma)
  -\beta_1(\phi)\lcd_\mu W^\mu-\beta_2(\phi)\lcd_\mu U^\mu-\beta_3(\phi)\lcd_\mu \bar T^\mu
  \\[2pt]
  &+\sum_{i=1}^{8}\gamma_i(\phi)\,J_i
  +\sum_{i=1}^{5}\omega_i(\phi)\,O_i
  \;-\tfrac12(\partial\phi)^2-V(\phi)\Big],
  \label{eq:actionPI}
\end{align}
with the parity-even basis 
\begin{alignat}{3}
  &J_1= \mathcal{T}_{\rho\mu\nu} \mathcal{T}^{\rho\mu\nu},\quad
  &&J_2= \mathcal{T}_{\rho\mu\nu} \mathcal{T}^{\mu\rho\nu},\quad
  &&J_3= \mathcal{Q}_{\rho\mu\nu} \mathcal{Q}^{\rho\mu\nu},\\
  &J_4= \mathcal{Q}_{\rho\mu\nu} \mathcal{Q}^{\mu\rho\nu},\quad
  &&J_5= \mathcal{Q}_{\rho\mu\nu} \mathcal{T}^{\mu\rho\nu},\quad
  &&J_6=W_\mu W^\mu,\\
  &J_7=W_\mu U^\mu,\quad
  &&J_8=U_\mu U^\mu, && \label{eq:Jbasis}
\end{alignat}
and the parity-odd basis 
\begin{alignat}{3}
  &O_1=\epsilon^{\mu\nu\rho\sigma} \mathcal{T}^\lambda{}_{\mu\nu} \mathcal{T}_{\lambda\rho\sigma},\quad
  &&O_2=\epsilon^{\mu\nu\rho\sigma} \mathcal{Q}_{\mu\rho\lambda} \mathcal{Q}_{\nu\sigma}{}^{\lambda},\quad
  &&O_3=\epsilon^{\mu\nu\rho\sigma} \mathcal{T}^\lambda{}_{\mu\nu} \mathcal{Q}_{\rho\sigma\lambda},\\
  &O_4=W_\mu \bar T^\mu,\quad
  &&O_5=U_\mu \bar T^\mu . && \label{eq:Obasis}
\end{alignat}
Note that  $\bar T_\mu \bar T^\mu$ does not appear as it is parity-even and lies in the span of $J_1$ and $J_2$, so it is not
independent.

Imposing projective invariance reduces the eleven-dimensional space of parity-even quadratic invariants to the eight-dimensional space spanned by $J_i$. Similarly, the six-dimensional parity-odd sector reduces to the five-dimensional space spanned by $O_i$.

At linear order in the distortion, the three independent projectively invariant vectors are $W^\mu$,
$U^\mu$ and $\bar T^\mu$. After integrating their Levi-Civita divergences $\cd_\mu W^\mu$, $\cd_\mu U^\mu$ and
$\cd_\mu \bar T^\mu$ by parts, the corresponding bulk contributions reduce to three operators proportional to the scalar gradient,
$\partial_\mu\phi W^\mu, \partial_\mu\phi U^\mu,\partial_\mu\phi\bar T^\mu$. The action can therefore be rewritten as 
\begin{equation}
\begin{aligned}
  \mathcal{S}=&\int\!\dd^4x\sqrt{-g}\;\Big[\;
  \tfrac12 F(\phi)\,R(\Gamma) +\partial_\mu\phi\,\big(\beta'_1(\phi)W^\mu+\beta'_2(\phi)U^\mu+\beta'_3(\phi)\bar T^\mu\big)
  \\[2pt]
  &+\sum_{i=1}^{8}\gamma_i(\phi)\,J_i
  +\sum_{i=1}^{5}\omega_i(\phi)\,O_i
  \;-\tfrac12(\partial\phi)^2-V(\phi)\Big]\\
  & -\int\!\dd^4x\sqrt{-g} \lcd_\mu( \beta_1 (\phi) W^\mu + \beta_2(\phi) U^\mu + \beta_3(\phi) \bar{T}^\mu)\,.
  \label{eq:action_2}
 \end{aligned}
\end{equation}
We stress that all the total derivatives in the linear distortion sector must be Levi-Civita ones. Since $\nabla_\mu V^\mu=\lcd_\mu V^\mu+N^\mu{}_{\mu\nu}V^\nu$ with $N^\mu{}_{\mu\nu}=T_\nu-\tfrac12 Q_\nu$, replacing a Levi-Civita divergence by an affine divergence would introduce an additional bulk term, which is not projectively invariant when $V^\mu$ is one of the invariant vectors considered here, as $\delta N^\mu{}_{\mu\nu}=\theta_\nu$.

In the action \eqref{eq:action_2}, the distortion enters through the Ricci scalar, $R(\Gamma)$, the quadratic invariants $J_i, O_i$ and the terms proportional to the $\beta'_i(\phi)$ functions. For generic choices of the coupling functions, the scalar gradient $\partial_\mu\phi$ sources a non-vanishing distortion, which feeds back into the scalar kinetic term\footnote{In theories where the connection enters only through the Ricci tensor instead, it is fixed, up to a projective term, to the Levi-Civita connection of an auxiliary metric~\cite{Afonso:2017bxr}.} (see sec. \ref{sec:integrate}). This mechanism modifies the Einstein-frame scalar potential and, consequently, the inflationary predictions.

\subsection{Absence of explicit Holst coupling}\label{sec:holst}
The action \eqref{eq:actionPI} contains no $c_H(\phi)\tilde R(\Gamma)$ term coupling the scalar field with the Holst invariant. This is not an
omission but a choice of basis, and no generality is lost. Since $\tilde R(g)\equiv0$ by the
first Bianchi identity, the Holst invariant is built \emph{entirely} out of the distortion,
\begin{equation}
  \tilde R(\Gamma)=-\lcd_\mu\bar T^\mu+\tfrac12 O_1+O_3-\tfrac23 O_4
  =-\lcd_\mu\bar T^\mu
  +\tfrac12\,\epsilon^{\mu\nu\rho\sigma}T^\lambda{}_{\mu\nu}T_{\lambda\rho\sigma}
  +\epsilon^{\mu\nu\rho\sigma}T^\lambda{}_{\mu\nu}Q_{\rho\sigma\lambda}.
  \label{eq:holstNY}
\end{equation}
Integrating the divergence by parts,
$-c_H(\phi)\,\cd_\mu \bar T^\mu\simeq \,c_H'(\phi)\,\partial_\mu\phi\,\bar T^\mu$, so that a Holst term
merely shifts the couplings $\beta'_3 (\phi), \omega_1(\phi), \omega_3(\phi), \omega_4(\phi)$. As these are already arbitrary functions of $\phi$, $c_H(\phi)$ can be set to zero.

By the same argument one could trade $F(\phi)R(\Gamma)$ for $F(\phi)\lcR(g)$ plus shifts of $\beta_1$, $\beta_2$ and $\gamma_{1,\dots,7}$. We retain it because it keeps the metric and Palatini limits manifest: the Palatini limit is $\beta_i=\gamma_i=\omega_i=0$, and the metric one is $\beta_1=-F$, $\beta_2=-F/2$, $\beta_3=0$. The action \eqref{eq:actionPI} therefore contains \emph{eighteen} functions\footnote{This set can be further reduced to 17 independent free functions after Weyl rescaling of the metric (see sec. \ref{sec:inflation}).},
\begin{equation}
  F(\phi),\quad V(\phi),\quad \beta'_{1,2,3}(\phi),\quad \gamma_{1\dots8}(\phi),\quad \omega_{1\dots5}(\phi),
\end{equation}
instead of the \emph{twenty-three} we would have without considering projective invariance. While this is a modest improvement, we will see in the next section that projective invariance further simplifies computations when it comes to integrating out the non-dynamical connection.

\section{Integrating out the distortion}
\label{sec:integrate}
Using \eqref{eq:XWU}, the decomposition \eqref{eq:Rdecomp} can be recast in a manifestly
projective-invariant form 
\begin{equation}\label{eq:RdecompJ}
  R(\Gamma)=\lcR(g)
  +\tfrac14 J_1+\tfrac12 J_2+\tfrac14 J_3-\tfrac12 J_4+J_5
  -\tfrac23 J_6-\tfrac23 J_7
  -\lcd_\mu\big(2W^\mu+U^\mu\big),
\end{equation}
with neither a $J_8=U_\mu U^\mu$ contribution nor any parity-odd contribution. Inserting \eqref{eq:RdecompJ}
into \eqref{eq:action_2} and integrating the divergence by parts yields 
\begin{equation}
  \mathcal{S}=\int\!\dd^4x\sqrt{-g}\,\Big[\tfrac12 F(\phi) \lcR(g)-\tfrac12(\partial\phi)^2-V(\phi)
  +\Lalg[T,Q;\phi]\Big]+\int\!\dd^4x\sqrt{-g}\;\cd_\mu B^\mu ,
  \label{eq:ibp}
\end{equation}
with $B^\mu=\tfrac12F(\phi)X^\mu - \beta_1(\phi) W^\mu - \beta_2(\phi) U^\mu -\beta_3(\phi) \bar{T}^\mu$ and
\begin{align}\label{eq:Lalg}
  \Lalg[T,Q;\phi]
  =\;&\Big(\gamma_1+\tfrac18F\Big)J_1
     +\Big(\gamma_2+\tfrac14F\Big)J_2
     +\Big(\gamma_3+\tfrac18F\Big)J_3
     +\Big(\gamma_4-\tfrac14F\Big)J_4
     \\[2pt]
    &+\Big(\gamma_5+\tfrac12F\Big)J_5
     +\Big(\gamma_6-\tfrac13F\Big)J_6
     +\Big(\gamma_7-\tfrac13F\Big)J_7
     +\gamma_8\,J_8\nonumber
     \\[2pt]
    &+\sum_{i=1}^{5}\omega_i\,O_i
     +\partial_\mu\phi\Big[\big(F'+\beta'_1\big)W^\mu
                          +\big(\tfrac12F'+\beta'_2\big)U^\mu
                          +\beta'_3\,\bar T^\mu\Big]\nonumber,
\end{align}
where we omitted the explicit $\phi$-dependence of the couplings to avoid cumbersome notation. The algebraic Lagrangian $\Lalg$ is at most quadratic in $(T,Q)$ and contains terms linear in the distortion sourced by $\partial_\mu\phi$. For generic choices of the couplings, a non-vanishing scalar gradient therefore sources the distortion. In the bulk Lagrangian, the distortion appears without derivatives; all derivatives of it are confined to the total-divergence term in
\eqref{eq:ibp}. It is therefore an auxiliary field, and its elimination reduces to an algebraic problem.

\subsection{The vector ansatz is the general solution}
The only source available to the connection is $\partial_\mu \phi$. Therefore, provided that the algebraic connection system is non-degenerate apart from the projective gauge direction, the sourced solution is linear in $\partial_\mu\phi$, with coefficients built from the metric $g_{\mu\nu}$, the Levi-Civita tensor $\epsilon_{\mu\nu\rho\sigma}$ and functions of $\phi$. The most general such configuration is

\begin{equation}
\begin{aligned}
 & T_{\rho\mu\nu}=t_1(\phi)\big(g_{\rho\mu}\partial_\nu\phi-g_{\rho\nu}\partial_\mu \phi\big)
                 +t_2(\phi)\,\epsilon_{\rho\mu\nu\sigma}\partial^\sigma\phi,\\&
  Q_{\rho\mu\nu}=q_1(\phi)\,\partial_\rho\phi g_{\mu\nu}
                 +q_2(\phi)\big(g_{\rho\mu}\partial_\nu\phi+g_{\rho\nu}\partial_\mu\phi\big).
  \label{eq:ansatz}
\end{aligned}
\end{equation}

 Notice that this form is exhaustive for the part of the connection sourced by a single scalar gradient: the mixed-symmetry part of the torsion cannot be constructed from $g_{\mu\nu}$ and a single vector, while $\epsilon_{\mu\nu\rho\sigma}$ cannot generate a non-vanishing structure symmetric in the last two indices of $Q_{\rho\mu\nu}$ using only $\partial_\mu\phi$. Thus, for a non-degenerate connection system, Eq.~\eqref{eq:ansatz} gives the general sourced solution up to a projective transformation \cite{Rigouzzo:2022yan}.

Since the solution \eqref{eq:ansatz} is built from the single vector $\partial_\mu\phi$, its projective gauge direction can be parametrized as $\theta_\mu=\lambda(\phi)\partial_\mu\phi$. Equation~\eqref{eq:projTQ} then gives the transformation of the four coefficient functions,
\begin{equation}
  t_1\to t_1-\lambda,\qquad q_1\to q_1-2\lambda, \qquad t_2\to t_2, \qquad q_2\to q_2,
\end{equation}
so the projective invariant combinations are given by 
\begin{equation}
  \sigma\equiv 2t_1-q_1,\qquad q_2,\qquad t_2 .
  \label{eq:invvars}
\end{equation}
These three combinations carry the projectively invariant physical content of the scalar-sourced connection.

The projective transformation therefore removes one of the four coefficients in Eq. (5.4) as a gauge degree of freedom, leaving the three invariant combinations $\sigma$, $q_2$ and $t_2$. To rewrite the solution directly in terms of these projectively invariant degrees of freedom, we substitute Eq. (5.4) into the invariant building blocks defined in Eqs. (3.3)–(3.5). The projective gauge mode then drops out: $t_1$ and $q_1$ either cancel out or enter only through the invariant combination $\sigma\equiv 2t_1-q_1$. Explicitly,

\begin{equation}
\begin{split}
  &W_\mu=\tfrac34\big(2\sigma-q_2\big)\partial_\mu\phi,\qquad
  U_\mu=\tfrac92\,q_2\,\partial_\mu\phi,\qquad
  \bar T_\mu=6\,t_2\,\partial_\mu\phi,\\[4pt]
  &\mathcal T_{\rho\mu\nu}=t_2\,\epsilon_{\rho\mu\nu\lambda}\partial^\lambda\phi,\qquad
  \mathcal Q_{\rho\mu\nu}=q_2\Big(g_{\rho\mu}\partial_\nu\phi+g_{\rho\nu}\partial_\mu\phi
                                  -\tfrac12 g_{\mu\nu}\partial_\rho\phi\Big).
\end{split}
  \label{eq:onansatz}
\end{equation}

A few remarks on this solution: $J_5$, $O_1$, $O_2$  vanish identically on \eqref{eq:ansatz} . Hence, only
fifteen of the eighteen functions affect the scalar sector: $F$, $\beta'_{1,2,3}$,
$\gamma_{1,2,3,4,6,7,8}$, $\omega_{3,4,5}$ and $V$. We stress that the coefficients of $J_5$, $O_1$ and $O_2$, namely $\gamma_5$, $\omega_1$ and $\omega_2$, are not redundant operators of the full theory. They vanish only on the connection configuration sourced by a single scalar gradient and can in general be probed by matter carrying additional hypermomentum components, for example a hyperfluid with non-vanishing shear hypermomentum.

\subsection{Solving for the affine connection}\label{sec:linear_system}
Substituting \eqref{eq:ansatz} into \eqref{eq:ibp} and writing $v=(\sigma,q_2,t_2)^{\rm T}$,
the algebraic sector takes the compact form 
\begin{equation}
  \Lalg\Big|_{\rm ansatz}
  =\Big[\,S^{\rm T}v+\tfrac12\,v^{\rm T}\!M\,v\,\Big](\partial\phi)^2 ,
  \label{eq:quadform}
\end{equation}
with $M=\tfrac32\hat M$, $S=\tfrac32\hat S$ and 
\begin{equation}
  \hat M=\begin{pmatrix}
    -\mathcal A & -\mathcal B & -\mathcal E\\
    -\mathcal B & \phantom{-}\mathcal C & -\mathcal H\\
    -\mathcal E & -\mathcal H & \phantom{-}\mathcal D
  \end{pmatrix},
  \qquad
  \hat S=\begin{pmatrix}
    S_1 \\[2pt]
    S_2  \\[2pt]
    S_3 
  \end{pmatrix}\equiv\begin{pmatrix}
    F'+\beta'_1 \\[2pt]
    F'-\tfrac12 \beta'_1+3\beta'_2 \\[2pt]
    4\,\beta'_3
  \end{pmatrix},
  \label{eq:Mhat}
\end{equation}
where $\mathcal A,\dots,\mathcal H$ are the six \emph{effective couplings} listed in
Appendix~\ref{app:coeff}, Eq.~\eqref{eq:effcoup}. All the dependence of the scalar sector on the eighteen coupling functions is carried by these six effective couplings together with the three components of $\hat S$, $S_{1,2,3}$, the non-minimal coupling and the scalar potential $F,V$, for a total of eleven independent functions.
 
The auxiliary variables in \eqref{eq:quadform} are the three projective invariants
$v=(\sigma,q_2,t_2)^{\rm T}$ of \eqref{eq:invvars}, i.e.\ the coefficients that parametrise the
general solution \eqref{eq:ansatz} modulo the projective gauge direction. They are the only
independent components of the distortion that a scalar gradient can excite, and it is with
respect to them
that
\eqref{eq:quadform} is to be varied. Every term in $\Lalg$ evaluated on \eqref{eq:ansatz} carries an overall factor of $(\partial\phi)^2$: the $J_i$ and $O_i$ are quadratic in the distortion and hence quadratic in both $v$ and $\partial_\mu\phi$, while the operators $\partial_\mu\phi\,W^\mu$, $\partial_\mu\phi\,U^\mu$ and $\partial_\mu\phi\,\bar T^\mu$ are linear in $v$ and again quadratic in $\partial_\mu\phi$. The factor $(\partial\phi)^2$ therefore pulls
out overall, and $\Lalg|_{\rm ansatz}$ collapses to the inhomogeneous quadratic polynomial
\eqref{eq:quadform} in the three variables $v$, with coefficients depending on $\phi$ alone.
In this language 
\begin{equation}
  S_i=\frac{1}{(\partial\phi)^2}\,
          \frac{\partial\,\Lalg\big|_{\rm ansatz}}{\partial v^{i}}\bigg|_{v=0},
  \qquad
  M_{ij}=\frac{1}{(\partial\phi)^2}\,
         \frac{\partial^{2}\,\Lalg\big|_{\rm ansatz}}{\partial v^{i}\,\partial v^{j}} ,
  \label{eq:MSdef}
\end{equation}
so $M$ is the Hessian of the algebraic Lagrangian with respect to $(\sigma,q_2,t_2)$, while
$S$ is its gradient at vanishing distortion. 

Before solving the connection field equation, a remark is in order. As anticipated in Sec.~\ref{sec:proj_inv}, projective invariance requires the dilation current to vanish. We define the hypermomentum of the non-curvature sector of \eqref{eq:action_2} in the standard way,
\begin{equation}\label{eq:hypdef}
  \Delta_\lambda{}^{\mu\nu}:=-\frac{2}{\sqrt{-g}}\frac{\delta\big(\sqrt{-g}\,\mathcal L_m\big)}{\delta\Gamma^\lambda{}_{\mu\nu}},
\end{equation}
with
\begin{equation}
  \mathcal L_m=-\frac{1}{2}(\partial\phi)^2-V(\phi)+\sum_{i=1}^{8}\gamma_i J_i+\sum_{i=1}^{5}\omega_i O_i
  +\partial_\mu\phi\big(\beta'_1 W^\mu+\beta'_2 U^\mu+\beta'_3\bar T^\mu\big).
\end{equation}
Under an arbitrary projective shift $\delta_\theta\Gamma^\lambda{}_{\mu\nu}=\delta^\lambda{}_\nu\,\theta_\mu$ one then has
\begin{equation}\label{eq:projvar}
  \delta_\theta\mathcal S_m=-\frac12\int\!\dd^4x\sqrt{-g}\;\Delta_\lambda{}^{\mu\lambda}\,\theta_\mu ,
\end{equation}
so projective invariance of the matter sector is equivalent to the vanishing of the dilation current, 
\begin{equation}\label{eq:dilzero}
  \Delta_\lambda{}^{\mu\lambda}=0\,.
\end{equation}
Since every operator in $\mathcal L_m$ is built from the invariant objects $\mathcal T_{\rho\mu\nu}$, $\mathcal Q_{\rho\mu\nu}$, $W_\mu$, $U_\mu$ and $\bar T^\mu$ of Sec.~\ref{sec:proj_inv}, $\delta_\theta\mathcal S_m=0$ for arbitrary $\theta_\mu(x)$. Hence \eqref{eq:dilzero} holds identically, i.e.\ off-shell and for arbitrary $T_{\rho\mu\nu}$ and $Q_{\rho\mu\nu}$: it is the Noether identity associated with projective invariance. This is not automatic for a generic non-minimally coupled matter sector, whose hypermomentum need not satisfy it.

Consider now the same identity, applied to the full algebraic Lagrangian $\Lalg$. In the non-degenerate sector considered below, the connection field equation $\delta(\sqrt{-g}\,\Lalg)/\delta\Gamma^\lambda{}_{\mu\nu}=0$ has three independent projections on the four-parameter subspace $(t_1,q_1,t_2,q_2)$, while the fourth vanishes identically as a consequence of projective invariance. Equivalently, on \eqref{eq:ansatz} $\Lalg$ depends on $t_1$ and $q_1$ only through $\sigma$. The undetermined direction is precisely the projective one, $t_1\to t_1-\lambda$, $q_1\to q_1-2\lambda$, and the equation closes on the three invariants $(\sigma,q_2,t_2)$ of \eqref{eq:invvars}.

The distortion field equation is the linear algebraic system 
\begin{equation}
  M\,v+S=0 ,
  \label{eq:eom}
\end{equation}
whose solution $v_\star=-M^{-1} S$ , substituted back into \eqref{eq:quadform}, yields
\begin{equation}
  \mathcal L\ \supset\ -\tfrac12\,(1+k(\phi))\,(\partial\phi)^2 ,
  \qquad
  k(\phi)=S^{\rm T}M^{-1}S
   =\tfrac32\,\hat S^{\rm T}\hat M^{-1}\hat S
   =-\frac{3}{2\hat\Delta}\;\hat S^{\rm T}\,\mathrm{adj}\big(\hat M\big)\,\hat S ,
  \label{eq:kdef}
\end{equation}
where $\hat\Delta\equiv-\det\hat M$, and in the following we restrict to the non-degenerate sector $\hat\Delta\neq0$. The $\mathrm{adj}\big(\hat M\big)$ is
\begin{equation}
  \mathrm{adj}\big(\hat M\big)=
  \begin{pmatrix}
    \mathcal C\mathcal D-\mathcal H^{2}
      & \mathcal B\mathcal D+\mathcal E\mathcal H
      & \mathcal B\mathcal H+\mathcal C\mathcal E\\[2pt]
    \mathcal B\mathcal D+\mathcal E\mathcal H
      & -\big(\mathcal A\mathcal D+\mathcal E^{2}\big)
      & \mathcal B\mathcal E-\mathcal A\mathcal H\\[2pt]
    \mathcal B\mathcal H+\mathcal C\mathcal E
      & \mathcal B\mathcal E-\mathcal A\mathcal H
      & -\big(\mathcal A\mathcal C+\mathcal B^{2}\big)
  \end{pmatrix}.
  \label{eq:adj}
\end{equation}
Explicitly, with $\hat S=(S_1,S_2,S_3)^{\rm T}$,
\begin{equation}
  \boxed{\;
  \begin{aligned}
  k(\phi)=\frac{3}{2\hat\Delta}\Big[\;
    &S_2^{\,2}\big(\mathcal A\mathcal D+\mathcal E^{2}\big)
    +S_3^{\,2}\big(\mathcal A\mathcal C+\mathcal B^{2}\big)
    -S_1^{\,2}\big(\mathcal C\mathcal D-\mathcal H^{2}\big)\\
    &-2S_1S_2\big(\mathcal B\mathcal D+\mathcal E\mathcal H\big)
     -2S_1S_3\big(\mathcal B\mathcal H+\mathcal C\mathcal E\big)
     +2S_2S_3\big(\mathcal A\mathcal H-\mathcal B\mathcal E\big)\Big]
  \end{aligned}\;}
  \label{eq:kclosed}
\end{equation}
with 
\begin{equation}
  \hat\Delta=\mathcal D\big(\mathcal A\mathcal C+\mathcal B^{2}\big)
            +2\,\mathcal B\mathcal E\mathcal H
            +\mathcal C\mathcal E^{2}
            -\mathcal A\mathcal H^{2} .
  \label{eq:Deltahat}
\end{equation}
The total scalar kinetic term is then $-\tfrac12\big(1+k(\phi)\big)(\partial\phi)^2$.

\section{Slow-roll inflation and the metric-affine $\xi$-attractors}
\label{sec:inflation}

\subsection{Einstein-frame action}
After dropping the boundary terms in \eqref{eq:ibp} and solving the linear system \eqref{eq:eom} the action reduces to:
\begin{equation}
    S= \int d^4x \sqrt{-g}\qty(\frac{1}{2}F(\phi)\mathring R(g) - \frac{1}{2}(1+k(\phi))(\partial\phi)^2 - V(\phi)),\,
\end{equation}
Assuming $F(\phi)>0$, we obtain the equivalent Einstein-frame formulation through the conformal transformation
\begin{equation}
    g_{\mu\nu}^E = \frac{F(\phi)}{M_P^2} g_{\mu\nu}, \,
\end{equation}
which yields the Einstein-frame equivalent action:
\begin{equation}\label{eq:einsteinframe_1}
    S= \int d^4x \sqrt{-g_E}\qty(\frac{M_P^2}{2}R(g_E) - \frac{1}{2}K(\phi)(\partial\phi)^2 - U(\phi)),\,
\end{equation}
where:
\begin{equation}\label{eq:kinetic}
    K(\phi) = M_P^2\qty(\frac{1+k(\phi)}{F(\phi)}+\frac{3}{2} \frac{F'(\phi)^2}{F(\phi)^2}), \, \qquad U(\phi) = M_P^4\frac{V(\phi)}{F(\phi)^2}.
\end{equation}
Finally, we introduce the canonically normalized Einstein-frame scalar field $\chi$ through
\begin{equation}\label{eq:canonical_scalar}
    \qty(\frac{\partial\chi}{\partial \phi})^2 = K(\phi).\,
\end{equation}
In terms of $\chi$, the action \eqref{eq:einsteinframe_1} takes the canonical form
\begin{equation}\label{eq:einsteinframe_2}
    S= \int d^4x \sqrt{-g_E}\qty(\frac{M_P^2}{2}R(g_E) - \frac{1}{2}(\partial\chi)^2 - U(\chi)).\,
\end{equation}
Notice that $U(\chi)$ does not in general admit a closed analytic expression, since the field redefinition \eqref{eq:canonical_scalar} cannot always be integrated and inverted analytically for arbitrary couplings. After canonical normalization, the effects of the independent connection are encoded in the field redefinition $\chi(\phi)$ and hence in the Einstein-frame potential $U(\chi)$. From here we can use the standard slow-roll prescription to compute the inflationary CMB observables.
Throughout this work, we restrict to a regular and positive kinetic function, $K(\phi)>0$, over the field range of interest. Cases in which $K(\phi)$ develops a pole will be investigated separately in Ref.~\cite{Dioguardi:2026_to_appear}.

\subsection{Slow-roll formalism}

In order to compute the CMB observables we define the following slow-roll parameters:
\begin{equation}\label{eq:sr}
    \epsilon_U (\chi) := \frac{M_P^2}{2}\qty(\frac{U'(\chi)}{U(\chi)})^2 ,\qquad \eta_U (\chi):= M_P^2\frac{U''(\chi)}{U(\chi)},
\end{equation}
where the $'$ denotes the derivative with respect to the argument $\chi$. The slow-roll condition is verified when $\epsilon_U(\chi),|\eta_U(\chi)|\ll 1$. These parameters can be directly related to observables in the CMB. The most direct comparison with the observations is made through the tensor-to-scalar ratio $r$, the scalar spectral index $n_s$, the amplitude of the scalar perturbations $A_s$. Their values depend on the number of $e$-folds $N_*$ between horizon exit and the end of inflation. The value of $N_*$ depends on the post-inflationary expansion history and, in particular, on the reheating phase. Since we do not specify a reheating mechanism, we consider the representative range $50<N_*<60$. At leading order in the slow-roll approximation, the inflationary observables and the number of $e$-folds are given in terms of $\epsilon_U, \eta_U$ by
\begin{align}\label{eq:sr_1}
r &= 16 \,\epsilon_U(\chi_*), \\ \label{eq:sr_2}
n_s &= 1 - 6\, \epsilon_U(\chi_*)+ 2 \,\eta_U (\chi_*),\\ \label{eq:sr_3}
A_s &= \frac{U(\chi_*)}{24\pi^2 M_P^4\,\epsilon_U(\chi_*)},\\ \label{eq:sr_4}
N_* &= \frac{1}{M_P^2} \int_{\chi_{\rm end}}^{\chi_*}\frac{U(\chi)}{U'(\chi)} d\chi,
\end{align}
where we defined $\chi_*$ as the value at which the pivot scale exits the horizon, and $\chi_{\rm end}$ as the value for which $\epsilon(\chi)_U = 1$ for the models studied in this work\footnote{Slow-roll can break down if $|\eta(\chi)| = 1$, however in most cases (such as the ones considered in this work) $\epsilon (\chi) = 1$ is reached first.}. Equations~\eqref{eq:sr_1}-\eqref{eq:sr_4} therefore allow us to compute the inflationary predictions directly from the Einstein-frame potential $U(\chi)$ appearing in \eqref{eq:einsteinframe_2}.

\subsection{A universal metric-affine $\alpha$-attractor}
\label{sec:general_f}

We now focus on studying the $\xi$-attractor models. Such models have been widely studied in the literature in the context of metric \cite{Kallosh:2013maa} and Palatini gravity \cite{Jarv:2017azx}. The main assumption of the model is that a single function $f(\phi)$ fixes both the non-minimal couplings and the potential and in particular
\begin{equation}
F(\phi)=M_P^2\big(1+\xi f(\phi)\big),\qquad
V(\phi)=M^4 f(\phi)^2\, ,  
\end{equation}
with $M$ a mass scale and $f(\phi)>0$ a dimensionless, monotonically increasing\footnote{Strictly speaking we only need $f'(\phi)>0$ in the region of field space where inflation takes place.} and unbounded function.
The model predicts different inflationary phenomenology depending on the underlying choice of the gravitational sector.
It is well-known that in the metric case this model predicts a Starobinsky attractor independently of the choice of $f(\phi)$.
On the other hand, in the Palatini framework, the attractor is lost: the predictions for $r, n_s$ become model dependent, i.e. they depend on the explicit form of $f(\phi)$, even though a simple relation between observables still exists, as recently shown in \cite{Dioguardi:2026dfw}.
Here we show for the first time that, if we extend the model to a generic MAG framework a universal attractor is recovered: an $\alpha$-attractor, set by a single parameter which depends on the distortion couplings. We regard this as the main structural result of the paper. 
We now show how this is achieved. First of all, let's extend the parametrization to the distortion sector, taking the quadratic distortion couplings to be proportional to $f(\phi)$ and the derivatives of the linear couplings to be proportional to $f'(\phi)$,
\begin{equation}
\begin{gathered}
  \gamma_i(\phi)=M_P^2\,\bar\gamma_i\, f(\phi),\qquad
  \omega_i(\phi)=M_P^2\,\bar \omega_i\, f(\phi),\qquad \beta'_i(\phi)=M_P^2\,\bar\beta_i\,f'(\phi),
\end{gathered}
  \label{eq:gf_couplings}
\end{equation}

Inserting \eqref{eq:gf_couplings} into \eqref{eq:effcoup} and \eqref{eq:Mhat}, every effective
coupling becomes linear in $f(\phi)$ and the source becomes proportional to $f'(\phi)$,
\begin{equation}
\begin{gathered}
  \mathcal A=M_P^2(1+\bar a\,f(\phi)),\qquad
  \mathcal B=M_P^2(1+\bar b\,f(\phi)),\qquad
  \mathcal C=M_P^2(2+\bar c\, f(\phi)),\\[2pt]
  \mathcal D=M_P^2(1+\bar d\,f(\phi)),\qquad
  \mathcal E=M_P^2\,\bar e\, f(\phi),\qquad
  \mathcal H=M_P^2\,\bar h\, f(\phi),\\[2pt]
  S_i=M_P^2\,f'(\phi)\,\bar s_i,\qquad 
\end{gathered}
  \label{eq:gf_effcoup}
\end{equation}
where $\bar a,\dots,\bar h$ and $\bar s_{1,2,3}$ are the constant combinations of $\xi,\bar \gamma_i,\bar \omega_i,\bar \beta_i$
given in \eqref{eq:abcdef}. 

Substituting \eqref{eq:gf_effcoup} into \eqref{eq:kdef} gives 
\begin{align}  \label{eq:gf_k}
  k(\phi)&= M_P^2 f'(\phi)^2 \tilde k(\phi)\equiv\,M_P^2\,f'(\phi)^2\frac{\kappa_1+\kappa_2 f(\phi)+\kappa_3 f(\phi)^2}{4\,\big(3+d_1 f(\phi)+d_2 f(\phi)^2+d_3 f(\phi)^3\big)},
\end{align}
where $\kappa_{1,2,3}$ and $d_{1,2,3}$ are the constants given in \eqref{eq:kappacoeffs}--\eqref{eq:dcoeffs}. Since $f(\phi)$ is monotonic it can be used as field variable and this allows us to write:
\begin{equation}
  U(f)=\frac{M^4 f^2}{(1+\xi f)^2},
  \qquad
  \qty(\frac{\dd\chi}{\dd f})^2
  =\frac{M_P^2\,\tilde k(f)}{1+\xi f}+\frac32\frac{M_P^2\,\xi^2}{(1+\xi f)^2}+\frac{1}{f'^2\,(1+\xi f)}.
  \label{eq:gf_chi}
\end{equation}
The potential $U(\chi)$ is defined parametrically by $U(f)$ and $\chi(f)$, so, at fixed values of the coupling constants, two models with different $f(\phi)$ differ only through the last term.
In particular, as we are now going to show explicitly, whenever the last term is negligible during inflation, $U(\chi)$ is independent of the choice of $f(\phi)$ and this restores the attractor behavior.

Consider now the regime $\xi f\gg1$. Assuming the dimensionless distortion couplings are at most of order $\xi$ and $d_3\neq0$, numerator and denominator of $\tilde k$ in \eqref{eq:gf_k} are dominated by their highest powers, and we have 
\begin{equation}
\tilde k\simeq\zeta/(4f)\,,\qquad \zeta\equiv\kappa_3/d_3,  
\end{equation}
where the $\zeta$ dependence on $\kappa_3, d_3$ includes both the distortion couplings and the conformal one $\xi$ (see    Appendix \ref{kphicoef}).
Defining
\begin{equation}
  \alpha_1(f,f')\equiv\frac{f}{M_P^2\,f'^2},
  \label{eq:gf_ell}
\end{equation}
with $f'$ expressed as a function of $f$, eq.~\eqref{eq:gf_chi} gives 
\begin{equation}
  \qty(\frac{\dd\chi}{\dd f})^2\simeq\frac{3M_P^2}{2f^2}\qty(\alpha_0+\frac{2\,\alpha_1(f,f')}{3\xi}),
  \qquad \alpha_0\equiv1+\frac{\zeta}{6\xi}.
  \label{eq:gf_alpha}
\end{equation}
%
Once $f$ is used as the field variable, the overall $f^{-2}$ dependence is common to all choices of $f(\phi)$. The dependence on the specific relation between $f$ and the original scalar field $\phi$ enters only through $\alpha_1(f,f')$.
Whenever $\alpha_1$ is constant or negligible compared to $\alpha_0$ over the inflationary range, one has
\begin{equation}
 \qty(\frac{\dd\chi}{\dd f})^2\simeq \frac{3\alpha M_P^2}{2f^2}\,,\qquad \alpha\equiv\alpha_0+\frac{2\alpha_1}{3\xi}\,.
\end{equation}
Choosing, with no loss of generality, the branch in which $\chi$ increases with $f$, we obtain
\begin{equation}\label{eq:chi_f}
\chi\simeq\sqrt{3\alpha/2}\,M_P\ln f,    
\end{equation}
up to an additive constant absorbed in a shift of $\chi$. Inverting Eq.~\eqref{eq:chi_f} gives $f$ as a function of the canonical field $\chi$, which can then be substituted into $U(f)$ \eqref{eq:gf_chi} to obtain the Einstein-frame potential $U(\chi)$ needed to compute the slow-roll parameters. In fact, recalling our choice of Jordan-frame potential as $V(\phi)=M^4f(\phi)^2$ and using again the large-field condition $\xi f\gg1$, and expanding the Einstein-frame potential to first order in $1/(\xi f)$, we obtain 
\begin{equation}
  U(\chi)\simeq\frac{M^4}{\xi^2}\qty(1-\frac{2}{\xi}\,
  e^{-\frac{2\chi}{\sqrt{6\alpha}\,M_P}}),
  \label{eq:gf_U}
\end{equation}
which is the large-field form of the $\alpha$-attractor potential \cite{Kallosh:2013yoa}.

From \eqref{eq:sr} and \eqref{eq:sr_4}, working consistently at leading order in the large-field expansion and neglecting the contribution of the end of inflation, 
\begin{equation}
  \epsilon_U\simeq\frac{4}{3\alpha\,(\xi f)^2},\qquad
  \eta_U\simeq-\frac{4}{3\alpha\,\xi f},\qquad
  N_*\simeq\frac{3\alpha}{4}\,\xi f_* ,
\end{equation}
with $f_*\equiv f(\phi_*)$ evaluated at horizon exit, so that, at leading order in the large-$N_*$ expansion,
\begin{equation}
  n_s\simeq1-\frac{2}{N_*},\qquad
  r\simeq\frac{12\,\alpha}{N_*^2},\qquad
  A_s\simeq\frac{M^4\,N_*^2}{18\pi^2\,\alpha\,\xi^2 M_P^4}.
  \label{eq:gf_obs}
\end{equation}
Since $\xi f_*\simeq4N_*/(3\alpha)$, the large-field expansion is justified at horizon exit
whenever $N_*\gg\alpha$, with corrections of relative order $\alpha/N_*$.\\
The dependence of $\alpha$ on $f(\phi)$ is entirely contained in $\alpha_1(f_*, f'_*)$:\\
\begin{itemize}
    \item for $f\simeq \frac{f_0\phi^2}{M_P^2}$, $\alpha_1=\frac{1}{4\,f_0}$ is constant and $\alpha=\alpha_0+1/(6f_0\xi)$,
    \item for a more general $f\propto (\frac{\phi}{M_p})^n$ one has $\alpha_1\propto f^{(2-n)/n}$ and $\frac{2\alpha_1}{3\xi}\propto N_*^{(2-n)/n}\,\xi^{-2/n}$.
\end{itemize}
At strong coupling, $\xi\gg N_*$, inflation takes place at $f_*\ll1$ and the correction vanishes for any $n$. Provided that $\alpha_0$ remains finite and nonzero, $\alpha \rightarrow \alpha_0$ and $U(\chi)$ is independent of $f(\phi)$, as for metric $\xi$-attractors.
\subsection{A note on the metric and Palatini formulation}
From eq.~\eqref{eq:gf_chi} we can see that $(\dd\chi/\dd f)^2$ receives three contributions: the first term, arising from
integrating out the distortion and fixed by the constant couplings alone, the conformal term,
arising from the conformal transformation of the metric Ricci scalar, and the third term, which
carries the functional form of $f(\phi)$ through $f'(\phi)$. In the Palatini limit the first two contributions cancel exactly, leaving only the $f$-dependent term. In the regime $\xi f\gg1$, universal predictions arise whenever the first two combined contributions dominate during inflation, i.e.\ for $\alpha_0\gg2\alpha_1/(3\xi)$.

Both metric and Palatini formulations are limits of the metric-affine one. Note that, by
\eqref{eq:abcdef}--\eqref{eq:dcoeffs}, $\alpha_0$ depends only on the ratios $\bar\gamma_i/\xi$,
$\bar\omega_i/\xi$ and $\bar\beta_i/\xi$. The metric formulation is recovered when the connection source vanishes, $\bar s_i=0$. In the non-degenerate sector this implies $v_\star=0$, so that the connection reduces to the Levi-Civita one up to the projective gauge mode, and consequently $k=0$. The conditions $\bar s_i=0$ correspond to $\bar\beta_1=-\xi$, $\bar\beta_2=-\xi/2$ and $\bar\beta_3=0$, independently of the values of $\bar\gamma_i$ and $\bar\omega_i$. These choice for the $\bar\beta_i$ cancel the connection source generated by the non-minimal coupling $F(\phi)$.

Only the conformal
and the $f$-dependent terms survive, $\alpha=1+2\alpha_1/(3\xi)$, and at strong coupling one
recovers the Starobinsky attractor for any $f(\phi)$. The value $\alpha_0\to1$ is also approached when the quadratic distortion couplings are large compared to $\xi$, so that $k(\phi)$ is suppressed and all distortion modes are frozen.
The Palatini case is recovered by switching off all distortion couplings,
$\bar\gamma_i=\bar\omega_i=\bar\beta_i=0$, for which $\zeta=-6\xi$ and $\alpha_0=0$: the first
term cancels the conformal one exactly, only the $f$-dependent term survives, and the attractor
is lost. Since $\alpha_0$ depends only on the ratios, the same happens whenever the distortion couplings become small compared to $\xi$. In particular, if $\bar\gamma_i$, $\bar\omega_i$ and $\bar\beta_i$ are kept fixed while $\xi\to\infty$, their ratios to $\xi$ vanish and the Palatini-like limit $\alpha_0\to0$ is approached.

We now consider instead the strong-coupling limit in which the ratios $\bar\gamma_i/\xi$, $\bar\omega_i/\xi$ and $\bar\beta_i/\xi$ are held fixed and finite. In this limit, the metric-affine couplings scale proportionally to $\xi$, the Einstein-frame potential $U(\chi)$ of the metric-affine model
\eqref{eq:gf_couplings} is independent of the choice of $f(\phi)$, and it approaches its plateau
as an $\alpha$-attractor with $\alpha=\alpha_0=1+\zeta/(6\xi)>0$. At leading order in $1/N_*$,
\begin{equation}
  n_s\simeq1-\frac{2}{N_*},\qquad r\simeq\frac{12\,\alpha_0}{N_*^2}.
\end{equation}
The predictions therefore coincide with those of metric $\xi$-attractors (Starobinsky,
$\alpha_0=1$) up to a single parameter $\alpha_0$, fixed by the distortion couplings, which
tunes $r$ at fixed $n_s$.

\noindent In other words, at strong coupling metric-affine gravity inflation behaves as metric gravity inflation plus
one parameter that controls the tensor-to-scalar ratio. The universality lost in the Palatini formulation is restored by the metric-affine operators, and it is lost again only on the hypersurface $\zeta=-6\xi$ ($\alpha_0=0$) of coupling space, which contains the Palatini model and where the dependence on $f(\phi)$ reappears as in the Palatini $\xi$-attractors
\cite{Dioguardi:2026dfw}.

\section{Numerical scan of the parameter space}
\label{sec:parameters}
\subsection{Minimal parametrization}
To render the whole analysis tractable for a numerical scan we now specify the couplings.
The functions $F,\gamma_i,\omega_i$ have mass dimension two, since $J_i,O_i\sim T^2$, while
the $\beta'_i$ have mass dimension one. For the numerical analysis, we first adopt the minimal dimensionless parametrization choice:
\begin{equation}
f(\phi) = \frac{\phi^2}{M_P^2}\,,
\end{equation}
which also fixes the potential to:
\begin{equation}
V=M^4{\frac{\phi^4}{M_P^4}}\equiv \lambda \phi^4\,.
\end{equation}
With this choice, we can compute $k(\phi)$ directly from \eqref{eq:gf_k}:
\begin{equation}
  k(\phi)=\frac{M_P^4 \kappa_1  \phi^2+ M_P^2 \kappa_2  \phi^4+\kappa_3\phi^6}
               {3 M_P^6+ M_P^4 d_1 \phi^2+ M_P^2 d_2 \phi^4+d_3\phi^6}\;,
  \label{eq:kpoly}
\end{equation}
with coefficients given in the appendix \eqref{eq:kappacoeffs}--\eqref{eq:dcoeffs}. Notice that the numerator and denominator are both cubic polynomials in $\phi^2$, and for generic couplings with $d_3\neq0$, one therefore has

\begin{equation}
  k\ \xrightarrow{\ \phi\gg\Mp\ }\ \zeta\equiv\frac{\kappa_3}{d_3}.
  \label{eq:kasympt}
\end{equation}
We stress again that in this work we are only considering the case $K(\phi)>0$ for every $\phi$. Cases in which $K(\phi)$ develops a pole will be investigated separately in Ref.~\cite{Dioguardi:2026_to_appear}.

In what follows, we study two well-behaved subsets of parameter space that illustrate the attractor behavior of the model.

\subsection{No hierarchy between the \texorpdfstring{$\bar\gamma_i$, $\bar\omega_i$, $\bar\beta_i$}{gamma, omega, beta} couplings}
\label{sec:no_hierarchy}

\begin{figure}[t!]%
    \centering
    {\includegraphics[width=1\textwidth]{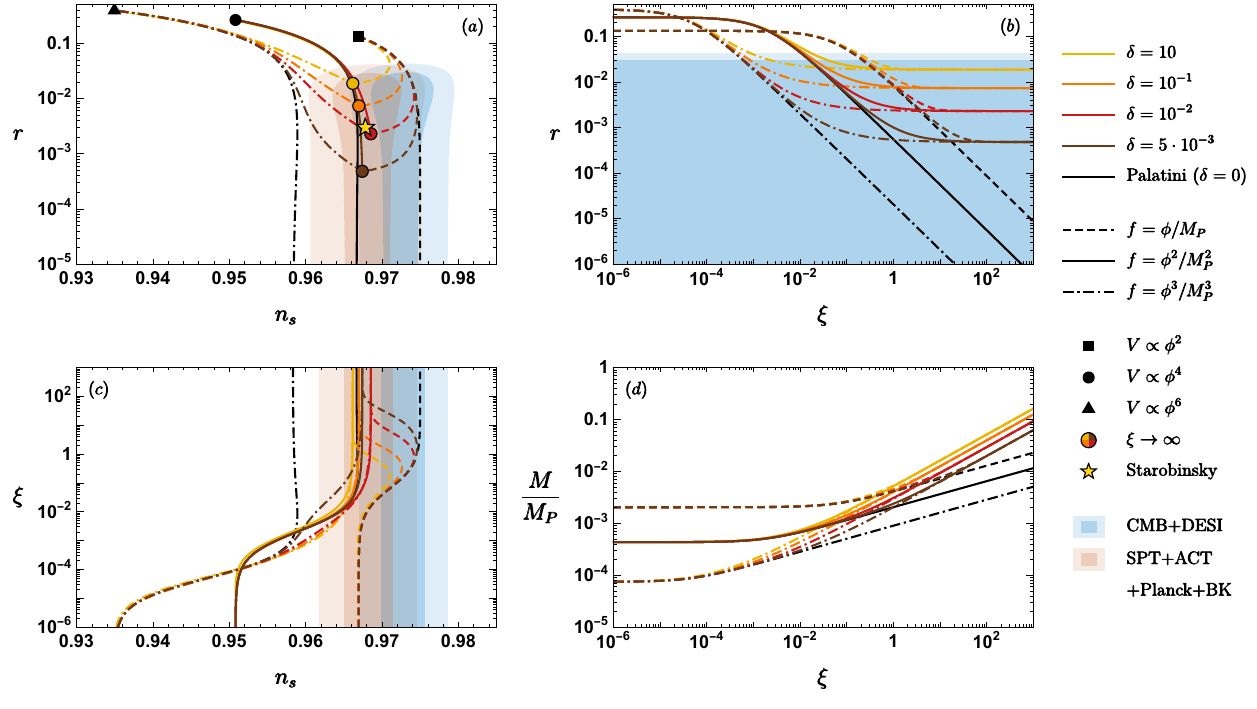}}%

   \caption{
   Slow-roll predictions at $N_*=60$ for $f(\phi)=\phi/\Mp$ (dashed), $\phi^2/\Mp^2$ (solid) and $\phi^3/\Mp^3$ (dot-dashed), with all distortion couplings scaling as $\delta\,\xi$, eq.~\eqref{eq:delta_scaling}, for $\delta=10$ (yellow), $10^{-1}$ (orange), $10^{-2}$ (red) and $5\cdot10^{-3}$ (brown). The black lines show the Palatini limit ($\delta=0$). Panel (a): $r$ vs $n_s$; (b): $r$ vs $\xi$; (c): $\xi$ vs $n_s$; (d): $M/\Mp$ vs $\xi$, with $M$ fixed by $A_s\simeq2.1\cdot10^{-9}$. The non-minimal Ricci coupling is taken in the interval $10^{-6}\le\xi\le10^{3}$. Filled square, circle and triangle show respectively the predictions of chaotic $\phi^2$, $\phi^4$ and $\phi^6$ inflation ($\xi\to0$); coloured discs are the predictions of the MAG $\alpha$-attractor ($\xi\to\infty$), while the yellow star is Starobinsky inflation. The shaded regions show the 68\% and 95\% CL from Planck, BICEP/Keck, ACT and SPT \cite{Balkenhol:2025wms} without (orange) and with (blue) DESI BAO \cite{DESI:2024mwx, DESI:2025zgx}.
   }
    \label{fig:model_1}
\end{figure}

To test the universality of the MAG $\alpha$-attractor derived in section~\ref{sec:general_f}, we now go beyond the minimal choice $f(\phi)=\phi^2/\Mp^2$ and consider
\begin{equation}
  f(\phi)=\qty(\frac{\phi}{\Mp})^{n},\qquad V(\phi)=M^4\qty(\frac{\phi}{\Mp})^{2n},\qquad n=1,2,3,
  \label{eq:fn_choice}
\end{equation}
i.e.\ quadratic, quartic and sextic Jordan-frame potentials, with all the couplings parametrized as in eq.~\eqref{eq:gf_couplings}. In this section the couplings of the linear, quadratic parity-even and quadratic parity-odd sectors are all positive, of the same order of magnitude, and scale together with $\xi$:
\begin{equation}
  \bar\gamma_i=\delta\,\xi\,\gamma^0_i,\qquad
  \bar\omega_i=\delta\,\xi\,\omega^0_i,\qquad
  \bar\beta_i=\delta\,\xi\,\beta^0_i,
  \label{eq:delta_scaling}
\end{equation}
where $\gamma^0_i,\omega^0_i,\beta^0_i$ are real numbers drawn uniformly in $[1,10]$ and kept fixed for all the curves\footnote{The prefactors are drawn randomly in order to avoid accidental cancellations in the effective couplings \eqref{eq:abcdef}. We checked that $K(\phi)>0$ for all $\phi$ along every curve.}. Here $\delta$ is a dimensionless parameter controlling the overall magnitude of the distortion couplings relative to the non-minimal coupling $\xi$. Since the ratios $\bar\gamma_i/\xi$, $\bar\omega_i/\xi$ and $\bar\beta_i/\xi$ do not depend on $\xi$, eq.~\eqref{eq:gf_alpha} predicts an attractor which depends on $\delta$ but neither on $\xi$ nor on $f(\phi)$. For our choice of the prefactors we obtain $\alpha_0=6.38,\,2.40,\,0.82,\,0.15$ for $\delta=10,\,10^{-1},\,10^{-2},\,5\cdot10^{-3}$, with $\alpha_0=1$ ($\zeta=0$) at $\delta\simeq1.1\cdot10^{-2}$, while $\delta=0$ corresponds to Palatini inflation ($\zeta=-6\xi$, $\alpha_0=0$).

The results are shown in Fig.~\ref{fig:model_1}. At small $\xi$ all the curves start from the predictions of the corresponding chaotic model, $V\propto\phi^{2n}$. The distortion corrections become relevant at smaller $\xi$ for larger $n$, since $f'(\phi_*)$ at horizon exit grows with $n$. As $\xi$ increases, for each $\delta$ the three choices of $f(\phi)$ merge onto the same point, which coincides with the $\xi\to\infty$ limit obtained by dropping the last, $f$-dependent, term of eq.~\eqref{eq:gf_chi} (coloured discs). The $f=\phi/\Mp$ curves are the last to converge, as expected from the size of the correction $2\alpha_1/(3\xi)$ in eq.~\eqref{eq:gf_alpha}: the three predictions for $r$ agree within $5\%$ for $\xi\gtrsim1$ when $\delta=10$, and only for $\xi\gtrsim60$ when $\delta=5\cdot10^{-3}$. At strong coupling we find $r\simeq1.9\cdot10^{-2},\,7.3\cdot10^{-3},\,2.3\cdot10^{-3},\,4.8\cdot10^{-4}$ for decreasing $\delta$, while $n_s\simeq0.966$--$0.969$ stays close to the $\alpha$-attractor value $1-2/N_*$, the small differences being subleading corrections in $1/N_*$. The metric-affine couplings therefore turn $r$ into a free parameter at fixed $n_s$: for $\delta\gtrsim1.1\cdot10^{-2}$ the tensor-to-scalar ratio is larger than the Starobinsky one, and within the reach of the next generation of CMB experiments, while for smaller $\delta$ it is suppressed.

The Palatini curves behave differently: there is no attractor, and the large-$\xi$ predictions depend on $f(\phi)$ \cite{Dioguardi:2026dfw}. For $f\propto\phi^n$ we find $r\to0$ and $n_s\to1-(n+2)/(2N_*)$, i.e.\ $n_s\simeq0.975,\,0.967,\,0.958$ for $n=1,2,3$ in agreement with \cite{Racioppi:2019jsp,Dioguardi:2026dfw}.

Panel (d) shows that also the mass scale of the potential, fixed by the amplitude of the scalar perturbations, is universal at strong coupling. Since $U(\chi)\to(M^4/\xi^2)\,u(\chi)$ with $u(\chi)$ independent of $f(\phi)$, the normalization of $A_s$ fixes $M/\sqrt\xi$ independently of the Jordan-frame potential: we find 
$M/(\Mp\sqrt\xi)\simeq5.1,\,3.9,\,2.9,\,2.0\times10^{-3}$ for decreasing $\delta$, in agreement within a few percent with 
\begin{equation}
M\simeq\Mp\,(18\pi^2\alpha_0A_s\,\xi^2/N_*^2)^{1/4}\,, 
\end{equation}
from eq.~\eqref{eq:gf_obs}. The dependence on $\delta$ enters only through $\alpha_0^{1/4}$. In the Palatini case, instead, $M$ depends on $f(\phi)$ for all values of $\xi$.

\subsection{Hierarchy between the quadratic and linear distortion couplings}
\label{sec:hierarchy}

\begin{figure}[t!]%
    \centering
    {\includegraphics[width=1\textwidth]{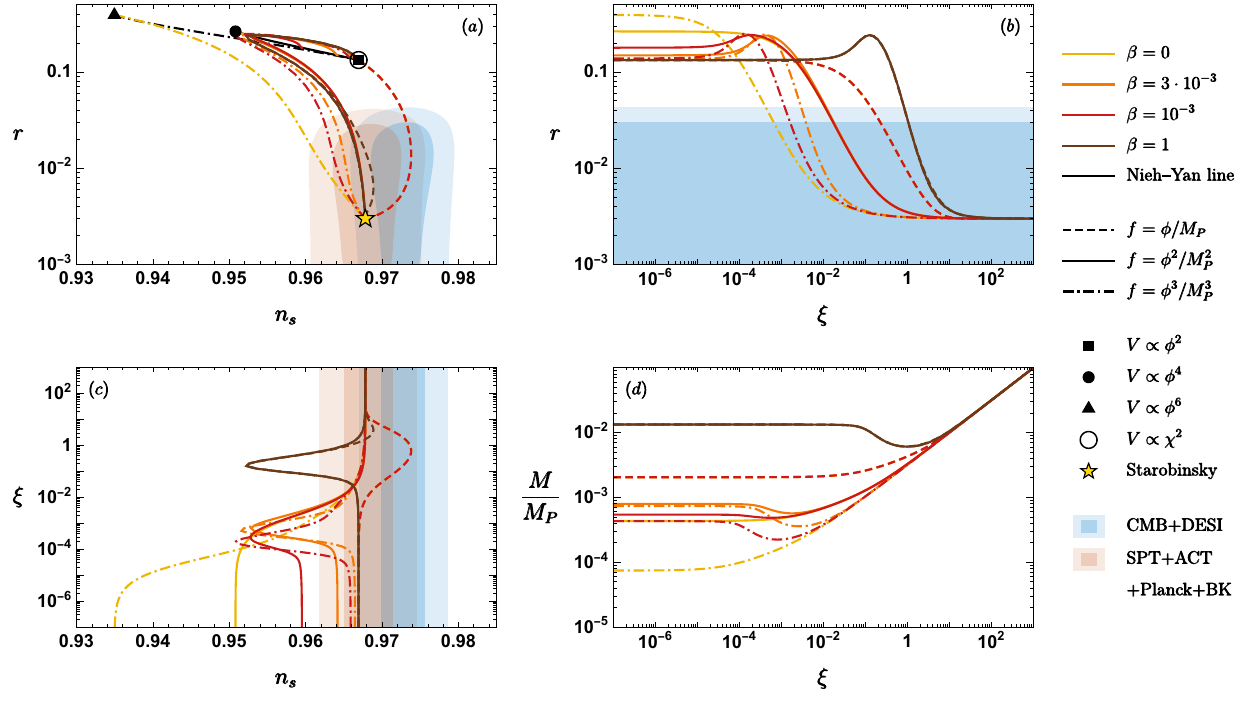}}%

   \caption{
   Same as Fig.~\ref{fig:model_1}, with the quadratic distortion couplings scaling as $\xi$ and the linear ones fixed by eq.~\eqref{eq:hierarchy_scaling}. Note that this time different colors show different values of $\beta$, in particular $\beta=0$ (yellow), $3\cdot10^{-3}$ (orange), $10^{-3}$ (red) and $1$ (brown); $10^{-7}\le\xi\le10^{3}$. The black lines show the Nieh--Yan-like limit $\xi=\bar\gamma_i=\bar\omega_i=0$ for $10^{-7}\le\beta\le10^{3}$, connecting chaotic $\phi^{2n}$ inflation (filled markers) to quadratic inflation (empty circle) for every $f(\phi)$. All the curves end on the Starobinsky point at large $\xi$ (yellow star). For $f=\phi/\Mp$ the curves with $\beta\le3\cdot10^{-3}$ overlap with the $\beta=0$ one, and for $\beta=1$ the three choices of $f(\phi)$ mostly overlap too.}
    \label{fig:model_2}
\end{figure}

In this section the quadratic couplings scale with $\xi$, while the linear ones are kept fixed:
\begin{equation}
  \bar\gamma_i=\xi\,\gamma^0_i,\qquad
  \bar\omega_i=\xi\,\omega^0_i,\qquad
  \bar\beta_i=\beta\,\beta^0_i,
  \label{eq:hierarchy_scaling}
\end{equation}
with $\beta=0,\,3\cdot10^{-3},\,10^{-3},\,1$ and $\gamma^0_i,\omega^0_i,\beta^0_i$ random real numbers with $|\gamma^0_i|,|\omega^0_i|,|\beta^0_i|\in[1,10]$\footnote{Contrary to the previous case, not all the prefactors can be taken positive. We checked that for our choices, $K(\phi)>0$ for all $\phi$ along every curve.}. At large $\xi$ the source $\hat S$ is dominated by $F'(\phi)$, i.e.\ $\bar s_{1,2}\simeq\xi\gg\bar s_3$ in eq.~\eqref{eq:abcdef}, and the linear couplings become irrelevant. At the same time the effective couplings \eqref{eq:abcdef} are, for our prefactors one to two orders of magnitude larger than $\xi$, so that $\zeta/(6\xi)\to0$, $\alpha_0 \rightarrow 1$ and all the curves end on the Starobinsky point, independently of $\beta$ and of $f(\phi)$.

In the opposite limit, $\xi=\bar\gamma_i=\bar\omega_i=0$, only the linear couplings survive and the model reduces to a Nieh--Yan-like one. In this case $\kappa_{2,3}=d_{1,2,3}=0$ in eqs.~\eqref{eq:kappacoeffs}--\eqref{eq:dcoeffs}, and eq.~\eqref{eq:gf_k} gives
\begin{equation}
  k(\phi)=\frac{\kappa_1}{12}\,\Mp^2 f'(\phi)^2,\qquad
  \kappa_1=6\qty(\bar s_2^2+3\bar s_3^2-2\bar s_1^2-2\bar s_1\bar s_2),
  \label{eq:k_NY}
\end{equation}
with $\bar s_i$ given by eq.~\eqref{eq:abcdef} at $\xi=0$. Since $F=\Mp^2$, the kinetic function is simply $K(\phi)=1+k(\phi)$, and for $\kappa_1>0$ and $k(\phi)\gg1$ the canonical field is
\begin{equation}
\chi(\phi)\simeq\sqrt{\kappa_1/12}\,\Mp f(\phi)\,,  
\end{equation} The potential then reads
\begin{equation}
  V(\phi)=M^4f(\phi)^2\simeq\frac{12\,M^4}{\kappa_1\Mp^2}\,\chi^2\,,
  \label{eq:V_NY}
\end{equation}
and the linear distortion couplings proportional to $f'(\phi)$ drive \emph{every} $\xi$-attractor potential towards quadratic chaotic inflation, independently of $f(\phi)$. This generalizes the quadratic attractor found for $f(\phi)=\phi^2/\Mp^2$ in \cite{Andrei:2026gmu}. The black lines in Fig.~\ref{fig:model_2} show this limit for $10^{-7}\le\beta\le10^{3}$: they connect the chaotic $\phi^{2n}$ predictions (filled markers) at small $\beta$ to the quadratic one (empty circle) at large $\beta$. For $f(\phi)=\phi/\Mp$, $k(\phi)$ is a constant which merely rescales $\phi$, and the line shrinks to a point.

For $\xi\to0$ the coloured curves therefore start on the black lines, at a point fixed by $\beta$: from the quadratic attractor for $\beta=1$, part-way along the black lines for $\beta=10^{-3},\,3\cdot10^{-3}$, and from chaotic $\phi^{2n}$ inflation for $\beta=0$. As $\xi$ increases, all of them flow to the Starobinsky point, which is reached (within $5\%$ in $r$) for $\xi\gtrsim0.6$--$10$ when $\beta\le3\cdot10^{-3}$ and for $\xi\gtrsim40$ when $\beta=1$. In the latter case the transition takes place at $\xi\simeq0.1$--$0.2$, where the $\xi$- and $\beta$-dependent parts of the source are comparable, and goes through a temporary increase of $r$ (up to $r\simeq0.24$) and decrease of $n_s$ (down to $n_s\simeq0.952$).

More generally, the predictions do not depend on $f(\phi)$ 
whenever the first two terms in Eq.~\eqref{eq:gf_chi}, arising from the distortion and the conformal transformation, dominate over the last, $f(\phi)$-dependent, one.

This happens at strong $\xi$-coupling (i.e. the Starobinsky attractor), but also at any $\xi$ when the linear couplings are large: for $\beta=1$ the three choices of $f(\phi)$ give the same predictions over the whole range of $\xi$, within $1.5\cdot10^{-3}$ in $n_s$, and the same mass scale $M\simeq1.3\cdot10^{-2}\,\Mp$ at small $\xi$. Conversely, for $f(\phi)=\phi/\Mp$ one has $f'(\phi)=1/\Mp$, and the purely linear contribution to $k(\phi)$ is a constant, $\kappa_1/12\lesssim2\cdot10^{-2}$ for $\beta\le3\cdot10^{-3}$: the corresponding curves cannot be distinguished from the $\beta=0$ one. At large $\xi$, finally, all the curves in panel (d) follow the single line $M\simeq3.1\cdot10^{-3}\sqrt\xi\,\Mp$ of the Starobinsky attractor.

\subsection{Random scan}
\label{sec:random_scan}

\begin{figure}[t!]%
    \centering
    {\includegraphics[width=1\textwidth]{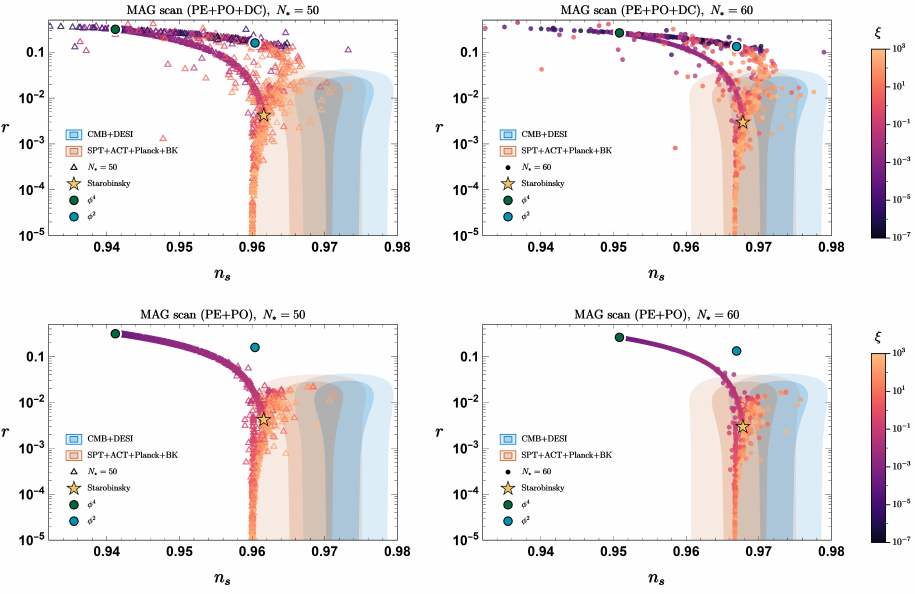}}%

   \caption{Slow-roll predictions of the random scan for $f(\phi)=\phi^2/\Mp^2$, i.e.\ $V(\phi)=\lambda\phi^4$, at $N_*=50$ (left) and $N_*=60$ (right). Top: all fourteen couplings switched on (PE+PO+DC: parity-even, parity-odd and linear distortion couplings); bottom: the same draws with $\bar\beta_i=0$ (PE+PO). Only ghost-free points are shown, coloured by $\xi$; points with $r<10^{-5}$ fall below the frame. The yellow star is Starobinsky inflation. The shaded regions show the 68\% and 95\% CL from Planck, BICEP/Keck, ACT and SPT \cite{Balkenhol:2025wms} without (orange) and with (blue) DESI BAO \cite{DESI:2024mwx, DESI:2025zgx}.}
    \label{fig:model_3}
\end{figure}

The two previous examples were built to isolate one attractor at a time. In order to see what a generic choice of the couplings predicts, we now scan the whole parameter space of the parametrization \eqref{eq:gf_couplings} with $f(\phi)=\phi^2/\Mp^2$, i.e.\ $V(\phi)=\lambda\phi^4$. The scalar sector depends on fourteen dimensionless couplings, $\xi$, $\bar\gamma_{1,2,3,4,6,7,8}$, $\bar\omega_{3,4,5}$ and $\bar\beta_{1,2,3}$ (we recall that $\bar\gamma_5$ and $\bar\omega_{1,2}$ drop out, see sec.~\ref{sec:integrate}), which enter $k(\phi)$ only through the ten combinations $\xi$, $\bar a,\dots,\bar h$ and $\bar s_{1,2,3}$ of eq.~\eqref{eq:abcdef}. Contrary to secs.~\ref{sec:no_hierarchy} and~\ref{sec:hierarchy}, we impose no relation among them and draw them independently: $\xi$ is taken positive (to ensure existence of the conformal transformation), each of the remaining thirteen couplings has a random sign, and all the absolute values are picked uniformly on a base-10 log-scale in the range $10^{-7}$--$10^{3}$. In this way every hierarchy between $\xi$, the quadratic and the linear distortion couplings is sampled.

Unfortunately not all the random picks are healthy. Since $K(0)=1$ and $F(\phi)K(\phi)/\Mp^2$ is the ratio of two quartic polynomials in $\phi^2$, a point is ghost-free, i.e.\ $K(\phi)>0$ for every $\phi$, if and only if neither polynomial has a positive root, which we check exactly by root counting. Only $\simeq24\%$ of the draws pass this test; for the remaining ones the kinetic term of the scalar field is negative in at least one region of field space. However, as long as $K(\phi)$ is well defined, inflation takes place and ends successfully. We keep $2\times10^3$ ghost-free points and compute their predictions numerically through eqs.~\eqref{eq:sr}--\eqref{eq:sr_4}; they are shown in the top panels of fig.~\ref{fig:model_3}.

In order to isolate the role of the linear distortion couplings, in the bottom panels we repeat the scan on the same underlying set of draws with $\bar\beta_i=0$. The source is then fixed to its Palatini value, $\bar s_1=\bar s_2=\xi$ and $\bar s_3=0$: the distortion is sourced by $F'(\phi)$ alone, and the quadratic couplings only determine how it responds. Since $\hat\Delta$, i.e.\ the denominator of eq.~\eqref{eq:kpoly}, does not depend on $\bar\beta_i$, all the difference between the two scans comes from the numerator of $k(\phi)$. Without the linear couplings the ghost-free fraction rises to $\simeq29\%$.

Even though there are fourteen a priori independent couplings, the predictions of the general model are quite simple. At leading order in $1/N_*$, all the attractors identified in the previous sections share the same spectral index, $n_s\simeq1-2/N_*$: the MAG $\alpha$-attractor, with $r\simeq12\alpha/N_*^2$ and $\alpha=\alpha_0+1/(6\xi)$ for $f(\phi)=\phi^2/\Mp^2$, and the Nieh--Yan-like quadratic attractor of eq.~\eqref{eq:V_NY} \cite{Andrei:2026gmu}, with $r\simeq8/N_*$. The bulk of the ghost-free points therefore lines up along a nearly vertical band, at $n_s\simeq0.96$ for $N_*=50$ and $n_s\simeq0.967$ for $N_*=60$, and the couplings mostly decide where along the band a point sits, i.e.\ the value of $r$. In the top panels, at large $\xi$ each point sits on the MAG $\alpha$-attractor with $\alpha_0$ fixed by its own ratios $\bar\gamma_i/\xi$, $\bar\omega_i/\xi$ and $\bar\beta_i/\xi$, which in the scan span many orders of magnitude, and so does $r$. Quadratic couplings large compared to $\xi$ freeze the distortion and drive $\alpha_0\to1$, i.e.\ to the Starobinsky point, while distortion couplings small compared to $\xi$ give the Palatini value $\alpha_0\to0$, for which $r\simeq2/(\xi N_*^2)$ drops below $10^{-5}$ for $\xi\gtrsim10^2$. Linear couplings large compared to $\xi$, instead, increase $\alpha_0$ and push the points up the band, towards the quadratic attractor, which is reached when they dominate the source. At small $\xi$, finally, the linear couplings still source the distortion, and the points spread between chaotic quartic inflation and the quadratic one, in the region spanned by the Nieh--Yan-like lines of fig.~\ref{fig:model_2}.

The bottom panels show what is left once the linear couplings are switched off. The Nieh--Yan-like attractor disappears by construction, since it is generated by the linear couplings alone, eq.~\eqref{eq:k_NY}. Moreover, for $\xi\to0$ the source vanishes altogether, $k(\phi)=\mathcal O(\xi^2)$, so all the weakly coupled points collapse onto chaotic $\lambda\phi^4$ inflation, independently of $\bar\gamma_i$ and $\bar\omega_i$. As $\xi$ increases, the points flow to the MAG $\alpha$-attractor, whose parameter is now set by the quadratic couplings alone:
\begin{equation}
  \alpha_0\big|_{\bar\beta_i=0}=1+\xi\,\frac{\bar d\,(\bar a-2\bar b-\bar c)+(\bar e-\bar h)^2}{d_3},
  \label{eq:alpha0_nolin}
\end{equation}
with $d_3$ given in eq.~\eqref{eq:dcoeffs}. Eq.~\eqref{eq:alpha0_nolin} gives the Palatini value $\alpha_0=0$ for $\bar\gamma_i=\bar\omega_i=0$ and the Starobinsky one, $\alpha_0\to1$, when the quadratic couplings are large compared to $\xi$; in general it depends on the ratios $\bar\gamma_i/\xi$ and $\bar\omega_i/\xi$. A tensor-to-scalar ratio different from the Starobinsky one at fixed $n_s$ is therefore obtained even without the linear couplings.

Finally, since all the attractors share $n_s\simeq1-2/N_*$, the agreement with the data is mostly a statement on $N_*$. For $N_*=60$ the band sits at $n_s\simeq0.967$--$0.968$, within the 68\% CL region of the CMB data alone and within the 95\% one once DESI is included, while for $N_*=50$ it lies at the edge of the 95\% CL region without DESI and outside it with DESI. Current data therefore favour a large number of $e$-folds, i.e.\ an efficient reheating (a high reheating temperature, for an equation of state $w<1/3$ during reheating). We remind that the density of points in the $(n_s,r)$ plane depends on how they are selected in the parameter space, and only the location of the attractors is a property of the model. However, as long as the couplings are picked in a wide enough range, the distribution of $(n_s,r)$ is qualitatively similar.

\section{Dropping projective invariance: the case of general MAG}\label{sec:MAG}

We discuss here the case of the general MAG action, obtained by dropping projective symmetry as a selection rule for the operators appearing in the action.
As we will see, the enlarged connection sector leads to a more involved computation and a less well-behaved parameter space, while the qualitative inflationary behavior remains unchanged.\\
The core result, however, still holds (with the hatted quantities defined in the following): the model predicts an $\alpha$-attractor in the regime $\hat\xi f(\phi)\gg1$, with exactly the same form of the projective invariant one and $\alpha=\hat\alpha_0+2\alpha_1(f,f')/(3\hat\xi)$, $\hat\alpha_0=1+\hat\zeta/(6\hat\xi)$, with $\hat \zeta \neq \zeta$ in general, given by the new asymptotic structure of $k(\phi)$. Whenever the $f(\phi)$-dependent correction is subdominant,
$\hat{\alpha}_0 \gg 2\alpha_1/(3\hat{\xi})$, at strong coupling, the predictions are therefore again independent of $f(\phi)$.

In the following subsections we derive the modified kinetic term obtained after integrating out the distortion in general MAG. We then show that the same $\alpha$-attractor emerges and briefly discuss the resulting parameter space and predictions in the $(n_s,r)$ plane.

\subsection{The general MAG action}
We start from the general MAG action with no propagating torsion and non-metricity, at most linear in curvature and quadratic in the distortion. The action is equivalent to the one studied in Ref.~\cite{Andrei:2026xad}, after fixing the scalar kinetic normalization to $B=1$ and translating between the conventions, as detailed in Appendix \ref{app:coeff_3}.

\begin{equation}
\begin{aligned}
S=\int\!\dd^4x\sqrt{-g}\;\Big[\;&\tfrac12 \hat{F}(\phi)R(\Gamma)  -\hat{\beta}_1(\phi)\lcd_\mu T^\mu-\hat{\beta}_2(\phi)\lcd_\mu Q^\mu-\hat{\beta}_3(\phi)\lcd_\mu \tilde Q^\mu -\hat{\beta}_4(\phi)\lcd_\mu \bar T^\mu \\
&+\sum_{i=1}^{11} \hat{\gamma}_i(\phi) \hat{J}_i+ \sum_{i=1}^{6} \hat{\omega}_i(\phi) \hat{O}_i-\tfrac{1}{2} (\partial\phi)^2-\hat{V}(\phi)\Big]
\end{aligned}\;
\label{eq:action}
\end{equation}

We can perform the integration by parts to get the bulk operators linear in the distortion and obtain
\begin{equation}
\begin{aligned}
S=\int\!\dd^4x\sqrt{-g}\;\Big[\;&\tfrac12 \hat{F}(\phi)R(\Gamma)
+\partial_\mu\phi\big(\hat{\beta}'_1(\phi)T^\mu +\hat{\beta}'_2(\phi)Q^\mu+ \hat{\beta}'_3(\phi)\Qt^\mu+ \hat{\beta}'_4(\phi) \bar T^\mu\big)\\
&+\sum_{i=1}^{11} \hat{\gamma}_i(\phi) \hat{J}_i+ \sum_{i=1}^{6} \hat{\omega}_i(\phi) \hat{O}_i-\tfrac{1}{2} (\partial\phi)^2-\hat{V}(\phi)\Big]\\
& -\int\!\dd^4x\sqrt{-g} \lcd_\mu( \hat{\beta}_1 (\phi)T^\mu + \hat{\beta}_2(\phi) Q^\mu + \hat{\beta}_3(\phi) \Qt^\mu + \hat{\beta}_4(\phi) \bar T^\mu)
\end{aligned}\;
\label{eq:action_ibp}
\end{equation}
with the parity-even basis (note: built from the \emph{full} $T,Q$, not from
their traceless parts).
\begin{alignat}{3}
&\hat{J}_1=T_{\rho\mu\nu}T^{\rho\mu\nu},&\quad
&\hat{J}_2=T_{\rho\mu\nu}T^{\mu\rho\nu},&\quad
&\hat{J}_3=Q_{\rho\mu\nu}Q^{\rho\mu\nu},\\
&\hat{J}_4=Q_{\rho\mu\nu}Q^{\mu\rho\nu},&\quad
&\hat{J}_5=Q_{\rho\mu\nu}T^{\mu\rho\nu},&\quad
&\hat{J}_6=T_\mu T^\mu,\nonumber\\
&\hat{J}_7=Q_\mu Q^\mu,&\quad
&\hat{J}_8=\Qt_\mu\Qt^\mu,&\quad
&\hat{J}_9=Q_\mu\Qt^\mu,\nonumber\\
&\hat{J}_{10}=Q_\mu T^\mu,&\quad
&\hat{J}_{11}=\Qt_\mu T^\mu,&&
\label{eq:Jbasis_2}
\end{alignat}
and the parity-odd basis 
\begin{alignat}{3}
&\hat{O}_1=\epsilon^{\mu\nu\rho\sigma}T^\lambda{}_{\mu\nu}T_{\lambda\rho\sigma},&\quad
&\hat{O}_2=\epsilon^{\mu\nu\rho\sigma}Q_{\mu\rho\lambda}Q_{\nu\sigma}{}^{\lambda},&\quad
&\hat{O}_3=\epsilon^{\mu\nu\rho\sigma}T^\lambda{}_{\mu\nu}Q_{\rho\sigma\lambda},\nonumber\\
&\hat{O}_4=\bar T_\mu T^\mu,&\quad
&\hat{O}_5=\bar T_\mu Q^\mu,&\quad
&\hat{O}_6=\bar T_\mu\Qt^\mu .
\label{eq:Obasis_2}
\end{alignat}
Upon imposing projective invariance and using the corresponding relations among the couplings, Eq.~\eqref{eq:action}  reduces to the projectively invariant action \eqref{eq:actionPI}.

\subsection{Modified kinetic term}

With the non-Riemannian Ricci scalar now given by 
\begin{equation}
\begin{gathered}
R(\Gamma)=\lcR(g)+\tfrac14 \hat{J}_1+\tfrac12 \hat{J}_2+\tfrac14 \hat{J}_3-\tfrac12 \hat{J}_4+ \hat{J}_5
-\hat{J}_6-\tfrac14 \hat{J}_7+\tfrac12 \hat{J}_9+\hat{J}_{10}-\hat{J}_{11}+\lcd_\mu X^\mu ,\\ \text{where} \hspace{1 cm}
X^\mu=Q^\mu-\Qt^\mu-2T^\mu,
\end{gathered}
\label{eq:Rdec}
\end{equation}
the action becomes, after integrating the divergence by parts,
\begin{equation}
S=\int\!\dd^4x\sqrt{-g}\Big[\tfrac12  \hat{F} \lcR(g)-\tfrac12(\partial\phi)^2- \hat{V}+\hat{\Lalg}\Big]+\int\!\dd^4x\sqrt{-g}\;\lcd_\mu \hat B^\mu,
\end{equation}
with $\hat B^\mu=\tfrac12\hat F X^\mu-\hat\beta_1T^\mu-\hat\beta_2Q^\mu-\hat\beta_3\Qt^\mu-\hat\beta_4\bar T^\mu$ and
\begin{equation}
\begin{aligned}
   \hat{\Lalg} \equiv& \sum_i \hat{\omega}_i \hat{O}_i
+\hat{J}_1 \left(\frac{\hat{F}}{8}+\hat{\gamma }_1\right)
+\hat{J}_2 \left(\frac{\hat{F}}{4}+\hat{\gamma }_2\right)
+\hat{J}_3 \left(\frac{\hat{F}}{8}+\hat{\gamma }_3\right)
+\hat{J}_4 \left(-\frac{\hat{F}}{4}+\hat{\gamma }_4\right)\\
&+\hat{J}_5 \left(\frac{\hat{F}}{2}+\hat{\gamma }_5\right)
+\hat{J}_6 \left(-\frac{\hat{F}}{2}+\hat{\gamma }_6\right)
+\hat{J}_7 \left(-\frac{\hat{F}}{8}+\hat{\gamma }_7\right)+\hat{J}_8 \hat{\gamma }_8+\hat{J}_9 \left(\frac{\hat{F}}{4}+\hat{\gamma }_9\right)\\
&+\hat{J}_{10} \left(\frac{\hat{F}}{2}+\hat{\gamma }_{10}\right)
+\hat{J}_{11} \left(-\frac{\hat{F}}{2}+\hat{\gamma }_{11}\right)+\partial _{\mu }\, \phi  \left(T^{\mu } \hat{\beta }_1'+Q^{\mu } \hat{\beta }_2'+\tilde{Q}^{\mu } \hat{\beta }_3'+\bar{T}^{\mu } \hat{\beta }_4'-\frac{1}{2} X^{\mu } \hat{F}'\right).
\end{aligned}
\end{equation}
Using the general solution \eqref{eq:ansatz} and writing $\hat{v}=(t_1,q_1,q_2,t_2)^{\rm T}$, following the same procedure of Sec.~\ref{sec:linear_system} we have
\begin{equation}
\hat{\Lalg}\big|_{\rm ansatz}=\Big[\,S^{\rm T} \hat{v}+\tfrac12\,\hat{v}^{\rm T}M \hat{v}\,\Big](\partial\phi)^2,
\qquad
S_i=\frac{1}{(\partial\phi)^2}\frac{\partial\hat{\Lalg}}{\partial \hat{v}^i}\bigg|_{\hat v=0},
\quad
M_{ij}=\frac{1}{(\partial\phi)^2}\frac{\partial^2\hat{\Lalg}}{\partial \hat{v}^i\partial \hat{v}^j},
\label{eq:quadform_2}
\end{equation}
and, with $M=\tfrac12\bar M$, $S=\tfrac12\bar S$,
\begin{equation}
\bar M=\begin{pmatrix}
-12\hat{\mathcal A}_1 & 6\hat{\mathcal A}_2 & -6\hat{\mathcal A}_3 & -12\hat{\mathcal B}_1\\
6\hat{\mathcal A}_2 & -\hat{\mathcal A}_4 & \hat{\mathcal A}_5 & 12\hat{\mathcal B}_2\\
-6\hat{\mathcal A}_3 & \hat{\mathcal A}_5 & \hat{\mathcal A}_6 & 12\hat{\mathcal B}_3\\
-12\hat{\mathcal B}_1 & 12\hat{\mathcal B}_2 & 12\hat{\mathcal B}_3 & 3\hat{\mathcal A}_7
\end{pmatrix},
\qquad
\bar S=\begin{pmatrix}
6\big(\hat F'+\hat{\beta}_1'\big)\\[2pt]
-3\hat F'+8\hat{\beta}_2'+2\hat{\beta}_3'\\[2pt]
3\hat F'+4\hat{\beta}_2'+10\hat{\beta}_3'\\[2pt]
12\,\hat{\beta}_4'
\end{pmatrix}\;
\label{eq:MS}
\end{equation}
where the ten \emph{effective couplings} are reported in the appendix \ref{effcoupgen}.
The original action \eqref{eq:action} contains $23$ scalar-dependent functions: $\hat F$, $\hat V$, the four $\hat\beta_i$, the eleven $\hat\gamma_i$ and the six $\hat\omega_i$. After evaluating the distortion sector on the general solution \eqref{eq:ansatz}, its dependence on these functions is encoded entirely in the ten effective couplings entering $\bar M$ and in the four components of $\bar S$ in \eqref{eq:MS}. Thus, the distortion-induced kinetic correction depends only on fourteen effective combinations of the original couplings. Notice that $\hat F$ also remains explicitly in the metric part of the action, while $\hat V$ does not enter the distortion sector.

The distortion field equation is again the linear system $M\hat v+S=0$, and substituting its solution back into \eqref{eq:quadform_2} gives
\begin{equation}
  k(\phi)=S^{\rm T}M^{-1}S=\tfrac12\,\bar S^{\rm T}\bar M^{-1}\bar S
  =\frac{\bar S^{\rm T}\,\mathrm{adj}\big(\bar M\big)\,\bar S}{2\det\bar M}\,,
  \label{eq:kdef_2}
\end{equation}
where we assume $\det\bar M\neq0$.

\subsection{$\alpha$-attractor again}
\label{sec:alpha_gen}

We now repeat the analysis of Sec.~\ref{sec:general_f}. We take $\hat F(\phi)=\Mp^2\big(1+\hat\xi f(\phi)\big)$, $\hat V(\phi)=M^4f(\phi)^2$ and, as in \eqref{eq:gf_couplings},
\begin{equation}
  \hat\gamma_i(\phi)=\Mp^2\,\hat{\bar\gamma}_i\,f(\phi),\qquad
  \hat\omega_i(\phi)=\Mp^2\,\hat{\bar\omega}_i\,f(\phi),\qquad
  \hat\beta'_i(\phi)=\Mp^2\,\hat{\bar\beta}_i\,f'(\phi),
  \label{eq:gf_couplings_2}
\end{equation}
with $\hat{\bar\gamma}_i$, $\hat{\bar\omega}_i$, $\hat{\bar\beta}_i$ dimensionless constants.

Considering $\bar S\equiv\Mp^2 f'(\phi)\,\hat{\bar s}$, the $\hat{\bar s}$ vector with the parametrization \eqref{eq:gf_couplings_2} becomes
\begin{equation}
\hat{\bar s}=\begin{pmatrix}
6(\hat{\xi}+\hat{\bar\beta}_1)\\
-3\hat{\xi}+8\hat{\bar\beta}_2+2\hat{\bar\beta}_3\\
3\hat{\xi}+4\hat{\bar\beta}_2+10\hat{\bar\beta}_3\\
12\hat{\bar\beta}_4
\end{pmatrix},
\label{eq:split}
\end{equation}
The effective couplings \eqref{eq:effcoup_2} of the matrix \eqref{eq:MS} also shall be written with the parametrization \eqref{eq:gf_couplings_2} and their expression is in appendix \ref{effcoupgen}. As in \eqref{eq:gf_effcoup}, they are all linear in $f(\phi)$.

We can now repeat the procedure followed in the projectively invariant case. Starting from the algebraic Lagrangian \eqref{eq:quadform_2}, we derive the distortion field equation, solve it for $\hat v$, and substitute the solution back into \eqref{eq:quadform_2}. Using the parametrization \eqref{eq:gf_couplings_2}, this gives the kinetic correction
\begin{equation}
  k(\phi)=\Mp^2 f'(\phi)^2\,\tilde k(\phi)\equiv\Mp^2 f'(\phi)^2\,
  \frac{\hat\kappa_1+\hat\kappa_2 f(\phi)+\hat\kappa_3 f(\phi)^2+\hat\kappa_4 f(\phi)^3}
       {4f(\phi)\big(\hat d_1+\hat d_2 f(\phi)+\hat d_3 f(\phi)^2+\hat d_4 f(\phi)^3\big)},
  \label{eq:kpoly_2}
\end{equation}
with the constants $\hat\kappa_n$, $\hat d_n$ given in appendix~\ref{kphicoeffgen}. This is the same structure as \eqref{eq:gf_k}, with numerator and denominator one degree higher since the four modes are independent.

In the regime $\hat\xi f\gg1$, assuming that the dimensionless distortion couplings are at most of order $\hat\xi$ and that $\hat d_4\neq0$, the highest powers of $f(\phi)$ dominate in \eqref{eq:kpoly_2}, giving
\begin{equation}
  \tilde k(\phi)\simeq\frac{\hat\zeta}{4f(\phi)},\qquad
  \hat\zeta\equiv\frac{\hat{\kappa}_4}{\hat{d}_4}\,,
  \label{eq:limits}
\end{equation}
so that, e.g., for $f(\phi)=\phi^2/\Mp^2$, this asymptotic form gives $k(\phi)\to\hat\zeta$. Since \eqref{eq:gf_chi} only requires $k=\Mp^2f'^2\tilde k$, the derivation of Sec.~\ref{sec:general_f} goes through unchanged, under the same assumption on the size of the couplings, with $\xi\to\hat\xi$ and $\zeta\to\hat\zeta$: eqs.~\eqref{eq:gf_alpha}--\eqref{eq:gf_obs} hold with
\begin{equation}
\alpha=\hat\alpha_0+\frac{2\,\alpha_1(f,f')}{3\hat\xi},\qquad
  \hat\alpha_0\equiv1+\frac{\hat\zeta}{6\hat\xi}.
  \label{eq:alpha_gen}
\end{equation}
Whenever the $f(\phi)$-dependent correction is subdominant,
$\hat\alpha_0\gg2\alpha_1/(3\hat\xi)$ over the inflationary range,
and at strong coupling, the potential $U(\chi)$ is again independent of $f(\phi)$. At leading order in $1/N_*$, the predictions are then $n_s\simeq1-2/N_*$ and $r\simeq12\hat\alpha_0/N_*^2$. As before, $\hat\alpha_0$ depends only on the ratios $\hat{\bar\gamma}_i/\hat\xi$, $\hat{\bar\omega}_i/\hat\xi$ and $\hat{\bar\beta}_i/\hat\xi$.

The metric limit is still $\bar S=0$, i.e.\ $\hat{\bar\beta}_1=-\hat\xi$, $\hat{\bar\beta}_2=-\hat{\bar\beta}_3=\hat\xi/2$, $\hat{\bar\beta}_4=0$, which gives $k\equiv0$ and the Starobinsky attractor is recovered at strong coupling.

The Palatini point, being projectively invariant, lies on $\det\bar M=0$, and therefore cannot be obtained by directly inverting $\bar M$ in \eqref{eq:kdef_2}, and is reached only as a limit, with $\hat\alpha_0\to0$.

\subsection{A note on the parameter space}

Also in this case we performed a random scan of the parameter space using parametrization \eqref{eq:gf_couplings_2} with $f(\phi)=\phi^2/\Mp^2$, i.e.\ $\hat V=\lambda\phi^4$. Including the potential normalization, the scalar sector depends on $22$ parameters: $\hat\xi$, the eleven $\hat{\bar\gamma}_i$, the five $\hat{\bar\omega}_i$ with $i\neq2$, the four $\hat{\bar\beta}_i$, and $\lambda$. Since $\lambda$ is fixed by the amplitude of the scalar perturbations, it is not independently scanned.

We selected the remaining 21 independent couplings considering again $\hat\xi>0$ and $\hat{\bar\gamma}_i,\hat{\bar\omega}_i,\hat{\bar\beta}_i$ all in the range $10^{-7}$ to $10^3$ in absolute value, with random signs. Once again the values have been picked using base-10 log-scale. The main difference with the projectively invariant case is a smaller ghost-free subspace yielding $\sim 14\%$ of well-behaved configurations. The remaining $86\%$ has at least one region of field space where $K(\phi)<0$. Apart from that, the $(n_s,r)$ plane predictions yield the same qualitative result as the projective invariant case and a plot would contain no new information. We conclude that the non-projective invariant operators do not reach any new inflationary phenomenology, hence the only role of non-projective invariant operators is to restrict the size of the healthy parameter space of the theory.

\section{Discussion and conclusions}
\label{sec:conclusions}

In this paper we studied single-field slow-roll inflation in metric-affine gravity. We considered the most general action containing operators at most linear in the curvature and quadratic in the distortion, without propagating torsion and non-metricity, and we promoted every coefficient to a function of the inflaton. When the couplings are constant and the scalar field is minimally coupled in the non-degenerate sector, the distortion has no source, the connection reduces to the Levi-Civita one and the theory is equivalent to General Relativity. This is no longer the case when the coefficients depend on the scalar field: the gradient $\partial_\mu\phi$ sources torsion and non-metricity, which in turn feed back into the scalar sector. The theory can still be recast as an Einstein--Hilbert term plus a minimally coupled canonical scalar, but this requires integrating out the non-dynamical distortion, a conformal transformation of the metric, with $F(\phi) >0$, and a scalar-field redefinition, with $K(\phi)$ well-defined, which together dramatically affect the inflationary predictions.

We first used projective invariance as a guiding principle to organise the coupling space. It reduces the twenty-three free functions of the general action to eighteen and contains the metric, Palatini, Holst and Nieh--Yan models as special cases, an explicit Holst coupling being redundant. Since the only available source is $\partial_\mu\phi$, the general solution of the connection field equation is a four-parameter family, one direction of which is pure projective gauge. Consistently with the vanishing of the dilation current, which holds identically as the Noether identity of projective symmetry, the connection equation collapses to a $3\times3$ linear system for the three invariants $(\sigma,q_2,t_2)$. Its solution gives the correction $k(\phi)$ to the scalar kinetic term in closed form, eq.~\eqref{eq:kclosed}, with denominator set by the determinant of the matrix $\hat M$. The scalar sector depends on the eighteen functions only through eleven quantities: the six effective couplings entering $\hat M$, the three source components $S_{1,2,3}$, together with the non-minimal coupling $F$ and the scalar potential $V$.

The main result of the paper concerns the $\xi$-attractor class, $F(\phi)=\Mp^2(1+\xi f(\phi))$ and $V(\phi)=M^4f(\phi)^2$, which we extended to metric-affine gravity by taking all distortion couplings proportional to $f(\phi)$.

In the metric formulation these models universally predict the Starobinsky attractor at strong coupling, while in the Palatini formulation the attractor is lost and the predictions depend on $f(\phi)$. We showed that the metric-affine framework restores universality aside from the Palatini limit $\alpha_0=0$): provided the kinetic function stays positive, at strong coupling the Einstein-frame potential becomes independent of $f(\phi)$ and approaches its plateau as an $\alpha$-attractor,
\begin{equation*}
  n_s\simeq1-\frac{2}{N_*},\qquad r\simeq\frac{12\,\alpha}{N_*^2},\qquad
  \alpha=\alpha_0+\frac{2\alpha_1(f,f')}{3\xi},\qquad \alpha_0=1+\frac{\zeta}{6\xi}.
\end{equation*}
The single constant $\zeta=\kappa_3/d_3$ summarizes the entire contribution of the linear and quadratic distortion couplings. 
The term $2\alpha_1(f,f')/(3\xi)$ carries the only dependence on $f(\phi)$: it equals $1/(6\xi)$ for $f=\phi^2/\Mp^2$ and vanishes at strong coupling, $\xi\gg N_*$, for any power law $f\propto\phi^n$. Starobinsky inflation, $\alpha_0=1$, is recovered when the source of the distortion vanishes, i.e.\ in the metric limit, or when the quadratic distortion couplings are large compared to $\xi$ and the distortion is frozen. Palatini inflation lies instead on the hypersurface $\zeta=-6\xi$, where $\alpha_0=0$ and the dependence on $f(\phi)$ reappears. At leading order in $1/N_*$ the metric-affine couplings do not shift $n_s$: they turn $r$ into a free parameter at fixed $n_s$. In the universal attractor regime, the normalization of $A_s$ fixes $M/\sqrt\xi$ independently of $f(\phi)$.

We verified this behavior numerically for $f=(\phi/\Mp)^n$, $n=1,2,3$. When all the distortion couplings scale together with $\xi$, the three choices of $f(\phi)$ merge at large $\xi$ onto the predicted attractor. For the representative choices shown in Fig.~\ref{fig:model_1}, we obtain $\alpha_0$ between $0.15$ and $6.38$ and $r$ between $\simeq5\cdot10^{-4}$ and $\simeq2\cdot10^{-2}$ at $n_s\simeq0.966$--$0.969$ for $N_*=60$, while the Palatini curves remain $f$-dependent. When only the quadratic couplings scale with $\xi$, every curve ends on the Starobinsky point. In the opposite limit, where only the linear couplings survive and generate a large kinetic correction, $k(\phi)\gg1$, every $\xi$-attractor potential is driven towards quadratic chaotic inflation, which generalises the Nieh--Yan-like attractor of \cite{Andrei:2026gmu} to arbitrary $f(\phi)$.

The attractor behaviour requires $K(\phi)$ to be well defined, and projective invariance alone is not sufficient to protect the model from ghosts. A random scan over the fourteen dimensionless couplings, spanning ten orders of magnitude with random signs, shows that only $\simeq24\%$ of the configurations are ghost-free ($\simeq29\%$ when the linear couplings are switched off). However, every healthy point produces successful inflation, and the predictions are much simpler than the size of the parameter space would suggest. The two attractors are the Nieh--Yan-like quadratic one, generated by the linear couplings, and the MAG $\alpha$-attractor found here, which contains the Starobinsky case for $\alpha_0=1$. The Palatini limit instead lies on the hypersurface $\alpha_0=0$, where the universal $\alpha$-attractor behaviour is lost.

Finally, we dropped projective invariance and repeated the analysis for the most general metric-affine action considered here. The scalar-sourced distortion then contains four independent modes and the connection equation becomes a $4\times4$ algebraic system. After integrating out the distortion, the kinetic correction is controlled by ten effective couplings and four source components, with only $\hat\omega_2$ dropping out. Despite this enlarged coupling space, the large-field structure remains the same as in the projectively invariant theory. In the regime $\hat\xi f(\phi)\gg1$, one finds $\tilde k\simeq\hat\zeta/(4f)$, with $\hat\zeta=\hat\kappa_4/\hat d_4$, and therefore the same $\alpha$-attractor structure with $\hat\alpha_0=1+\hat\zeta/(6\hat\xi)$. Whenever the $f(\phi)$-dependent correction is subdominant, the Einstein-frame potential again becomes independent of the specific choice of $f(\phi)$ and approaches the universal $\alpha$-attractor predictions, $n_s\simeq1-2/N_*$ and $r\simeq12\hat\alpha_0/N_*^2$. The metric limit reproduces the Starobinsky attractor, while the Palatini point is recovered as the singular projectively invariant limit $\hat\alpha_0\to0$.
This shows that the metric-affine $\alpha$-attractor is not a consequence of imposing projective symmetry, but persists in the general MAG theory considered here. The additional non-projectively invariant operators enlarge the connection sector and substantially reduce the healthy region of parameter space, with the ghost-free fraction decreasing from about $24\%$ in the projectively invariant scan to about $14\%$ in the general case, but they do not introduce qualitatively new inflationary predictions.

Throughout this work we assumed a regular kinetic function $K(\phi)>0$, for every $\phi$. The pole structure of $K(\phi)$, inherited from the zeros of $\det\hat M$, is however very rich, and its consequences for inflation will be discussed in a separate work \cite{Dioguardi:2026_to_appear}. On the observational side, the metric-affine $\alpha$-attractor shares the spectral index of Starobinsky inflation and differs from it only through $r$, so the next generation of CMB experiments \cite{SimonsObservatory:2018koc, LiteBIRD:2020khw} will directly probe a large part of its parameter space.

\appendix

\section{Coefficients for the projectively invariant action}
\label{app:coeff}

\subsection{Effective couplings} 
The six effective couplings entering $\hat M$ in \eqref{eq:Mhat} are
\begin{equation}
\begin{aligned}
  \mathcal A&=F-3\gamma_6, \\
  \mathcal B&=F+\tfrac32\gamma_6-\tfrac92\gamma_7, \\
  \mathcal C&=2F+12\gamma_3+3\gamma_4+\tfrac34\gamma_6-\tfrac92\gamma_7+27\gamma_8, \\
  \mathcal D&=F-8(\gamma_1-\gamma_2), \\
  \mathcal E&=-6\omega_4, \\
  \mathcal H&=6\omega_3+3\omega_4-18\omega_5 .
\end{aligned}
\label{eq:effcoup}
\end{equation}
The functions $\gamma_5$, $\omega_1$ and $\omega_2$ do not appear. Moreover, $\gamma_{1,2}$ enter
only through $\gamma_1-\gamma_2$, $\gamma_{3,4,8}$ only through $4\gamma_3+\gamma_4+9\gamma_8$, and
$\omega_{3,5}$ only through $\omega_3-3\omega_5$. A Holst coupling $c_H(\phi)\tilde R(\Gamma)$,
absorbed as in Sec.~\ref{sec:holst} through $\beta'_3\to\beta'_3+c_H'$,
$\omega_1\to\omega_1+\tfrac12c_H$, $\omega_3\to\omega_3+c_H$ and
$\omega_4\to\omega_4-\tfrac23c_H$, shifts both $\mathcal E$ and $\mathcal H$ by $4c_H$.
\subsection{Dimensionless coefficients}
For the $\xi$-attractor parametrization \eqref{eq:gf_couplings}, i.e.\ $F=M_P^2(1+\xi f)$,
$\gamma_i=M_P^2\bar\gamma_i f$, $\omega_i=M_P^2\bar\omega_i f$ and $\beta'_i=M_P^2\bar\beta_i f'$,
the effective couplings and the source take the form \eqref{eq:gf_effcoup} with
\begin{equation}
\begin{aligned}
  \bar a&=\xi-3\bar\gamma_6, \\
  \bar b&=\xi+\tfrac32\bar\gamma_6-\tfrac92\bar\gamma_7, \\
  \bar c&=2\xi+12\bar\gamma_3+3\bar\gamma_4+\tfrac34\bar\gamma_6-\tfrac92\bar\gamma_7+27\bar\gamma_8, \\
  \bar d&=\xi-8(\bar\gamma_1-\bar\gamma_2), \\
  \bar e&=-6\bar\omega_4, \\
  \bar h&=6\bar\omega_3+3\bar\omega_4-18\bar\omega_5 ,
\end{aligned}
\qquad\qquad
\begin{aligned}
  \bar s_1&=\xi+\bar\beta_1,\\
  \bar s_2&=\xi-\tfrac12\bar\beta_1+3\bar\beta_2,\\
  \bar s_3&=4\bar\beta_3 .
\end{aligned}
\label{eq:abcdef}
\end{equation}
For $f=\phi^2/M_P^2$, used in Sec.~\ref{sec:parameters}, the parametrization reduces to
$\gamma_i=\bar\gamma_i\phi^2$, $\omega_i=\bar\omega_i\phi^2$ and $\beta'_i=2\bar\beta_i\phi$.
\subsection{Coefficients of $k(\phi)$ \eqref{eq:gf_k}}\label{kphicoef}
With the shorthands
\begin{equation}
  P_1=2\bar a+2\bar b+\bar c,\qquad
  P_2=\bar a\bar c+\bar b^2,\qquad
  Z=2\bar e\bar h+2\bar e^2-\bar h^2 ,
\end{equation}
the numerator coefficients of \eqref{eq:gf_k} are 
\begin{equation}
\begin{aligned}
  \kappa_1&=6\Big[\bar s_2^2+3\bar s_3^2-2\bar s_1^2-2\bar s_1\bar s_2\Big], \\[3pt]
  \kappa_2&=6\Big[\bar s_2^2(\bar a+\bar d)+\bar s_3^2P_1-\bar s_1^2(\bar c+2\bar d)
        -2\bar s_1\bar s_2(\bar b+\bar d)-2\bar s_1\bar s_3(\bar h+2\bar e)
        +2\bar s_2\bar s_3(\bar h-\bar e)\Big], \\[3pt]
  \kappa_3&=6\Big[\bar s_2^2(\bar a\bar d+\bar e^2)+\bar s_3^2P_2-\bar s_1^2(\bar c\bar d-\bar h^2)
        -2\bar s_1\bar s_2(\bar b\bar d+\bar e\bar h)-2\bar s_1\bar s_3(\bar b\bar h+\bar c\bar e)
        +2\bar s_2\bar s_3(\bar a\bar h-\bar b\bar e)\Big],
\end{aligned}
\label{eq:kappacoeffs}
\end{equation}
and the denominator coefficients are 
\begin{equation}
\begin{aligned}
  d_1&=P_1+3\bar d, \\
  d_2&=P_2+\bar d\,P_1+Z, \\
  d_3&=\bar d\,P_2+2\bar b\bar e\bar h+\bar c\bar e^2-\bar a\bar h^2 .
\end{aligned}
\label{eq:dcoeffs}
\end{equation}
They follow from \eqref{eq:kclosed} and \eqref{eq:Deltahat}, with
$\hat\Delta=M_P^6\big(3+d_1f+d_2f^2+d_3f^3\big)$. In the regime $\xi f\gg1$ (assuming that $d_3\neq0$) one has
$\tilde k\simeq\zeta/(4f)$ with $\zeta\equiv\kappa_3/d_3$. For $f=\phi^2/M_P^2$,
\eqref{eq:gf_k} reduces to \eqref{eq:kpoly} with the same coefficients, so that
$k\simeq\kappa_1\phi^2/(3M_P^2)$ for $\xi\phi^2\ll M_P^2$ and $k\simeq\zeta$ for
$\xi\phi^2\gg M_P^2$.
\subsection{Consistency checks}
For general coupling functions, \eqref{eq:kclosed} reproduces the known limits 
\begin{alignat}{2}
  &\beta'_i=\gamma_i=\omega_i=0: \qquad
  &&k=-\frac{3F'^2}{2F}, \nonumber\\[4pt]
  &\beta'_3=-c_{NY}',\ \text{all others zero}: \qquad
  &&k=\frac{48c_{NY}'^2-3F'^2}{2F}, \nonumber\\[4pt]
  &\beta'_3=c_H',\ \mathcal E=\mathcal H=4c_H,\ \text{all others zero}: \qquad
  &&k=-\frac{3\big[32c_Hc_H'F'+F\big(F'^2-16c_H'^2\big)\big]}{2\big(F^2+16c_H^2\big)},
  \nonumber
\end{alignat}
corresponding to Palatini inflation, to a Nieh--Yan coupling $c_{NY}(\phi)\,\lcd_\mu\bar T^\mu$ and to a Holst coupling $c_H(\phi)\tilde R(\Gamma)$, respectively. The Jordan frame distortion-induced correction k is $k\equiv0$ whenever $\hat S=0$, i.e.\ for $\beta'_1=-F'$, $\beta'_2=-F'/2$ and $\beta'_3=0$. In this way, in the non-degenerate sector, we recover the metric formulation. In the Palatini case, the Einstein-frame kinetic terms \eqref{eq:kinetic} gives $K=M_P^2/F$.

For the $\xi$-attractor parametrization,
the Palatini and metric limits can also be checked directly at the level of the dimensionless coefficients. In the Palatini case, the dimensionless coefficients of the Jordan frame kinetic terms are:
$\bar\gamma_i=\bar\omega_i=\bar\beta_i=0$ which gives $\bar a=\bar b=\bar d=\xi$, $\bar c=2\xi$,
$\bar e=\bar h=0$, $\bar s_1=\bar s_2=\xi$ and $\bar s_3=0$, hence 
\begin{equation}
  \kappa_1=-18\xi^2,\quad \kappa_2=-36\xi^3,\quad \kappa_3=-18\xi^4,\qquad
  d_1=9\xi,\quad d_2=9\xi^2,\quad d_3=3\xi^3,
\end{equation}
so that we obtain $\tilde k=-\tfrac32\,\xi^2/(1+\xi f)$ and $\zeta=-6\xi$. In the $\xi$-attractor parametrization, the condition for recovering the metric case is $\bar s_i=0$, i.e.\ $\bar\beta_1=-\xi$, $\bar\beta_2=-\xi/2$ and $\bar\beta_3=0$. Since each $\kappa_i$ is quadratic in the source coefficients $\bar s_i$, one has $\kappa_i=0$ and therefore $k\equiv0$, independently of the quadratic distortion couplings.

\section{Coefficients for the general action}\label{app:coeff_2}
\subsection{Effective couplings}\label{effcoupgen}
The ten effective couplings entering the matrix \eqref{eq:MS} are
\begin{equation}
\begin{aligned}
\hat{\mathcal A}_1&=\hat{F}-2\hat{\gamma}_1-\hat{\gamma}_2-3\hat{\gamma}_6,\\
\hat{\mathcal A}_2&=\hat{F}-\hat{\gamma}_5+4\hat{\gamma}_{10}+\hat{\gamma}_{11},\\
\hat{\mathcal A}_3&=\hat{F}-\hat{\gamma}_5-2\hat{\gamma}_{10}-5\hat{\gamma}_{11},\\
\hat{\mathcal A}_4&=3\hat{F}-16\hat{\gamma}_3-4\hat{\gamma}_4-64\hat{\gamma}_7-4\hat{\gamma}_8-16\hat{\gamma}_9,\\
\hat{\mathcal A}_5&=3\hat{F}+8\hat{\gamma}_3+20\hat{\gamma}_4+32\hat{\gamma}_7+20\hat{\gamma}_8+44\hat{\gamma}_9,
\end{aligned}
\qquad
\begin{aligned}
\hat{\mathcal A}_6&=6\hat{F}+40\hat{\gamma}_3+28\hat{\gamma}_4+16\hat{\gamma}_7+100\hat{\gamma}_8+40\hat{\gamma}_9,\\
\hat{\mathcal A}_7&=\hat{F}-8\hat{\gamma}_1+8\hat{\gamma}_2,\\
\hat{\mathcal B}_1&=4\hat{\omega}_1-3\hat{\omega}_4,\\
\hat{\mathcal B}_2&=\hat{\omega}_3+4\hat{\omega}_5+\hat{\omega}_6,\\
\hat{\mathcal B}_3&=-\hat{\omega}_3+2\hat{\omega}_5+5\hat{\omega}_6 .
\end{aligned}
\label{eq:effcoup_2}
\end{equation}

The effective couplings of the matrix \eqref{eq:MS} with the parametrization \eqref{eq:gf_couplings_2} are
\begin{equation}
\begin{aligned}
\hat{\mathcal A}_1&=\Mp^2\big(1+\hat{\bar a}_1 f\big),\\
\hat{\mathcal A}_2&=\Mp^2\big(1+\hat{\bar a}_2 f\big),\\
\hat{\mathcal A}_3&=\Mp^2\big(1+\hat{\bar a}_3 f\big),\\
\hat{\mathcal A}_4&=\Mp^2\big(3+\hat{\bar a}_4 f\big),\\
\hat{\mathcal A}_5&=\Mp^2\big(3+\hat{\bar a}_5 f\big),
\end{aligned}
\qquad
\begin{aligned}
\hat{\mathcal A}_6&=\Mp^2\big(6+\hat{\bar a}_6 f\big),\\
\hat{\mathcal A}_7&=\Mp^2\big(1+\hat{\bar a}_7 f\big),\\
\hat{\mathcal B}_1&=\Mp^2\,\hat{\bar b}_1 f,\\
\hat{\mathcal B}_2&=\Mp^2\,\hat{\bar b}_2 f,\\
\hat{\mathcal B}_3&=\Mp^2\,\hat{\bar b}_3 f.
\end{aligned}
\label{eq:effcoup_minimal}
\end{equation}
where $f\equiv f(\phi)$ and
\begin{equation}
\begin{aligned}
\hat{\bar a}_1&=\hat\xi-2\hat{\bar\gamma}_1-\hat{\bar\gamma}_2-3\hat{\bar\gamma}_6,\\
\hat{\bar a}_2&=\hat\xi-\hat{\bar\gamma}_5+4\hat{\bar\gamma}_{10}+\hat{\bar\gamma}_{11},\\
\hat{\bar a}_3&=\hat\xi-\hat{\bar\gamma}_5-2\hat{\bar\gamma}_{10}-5\hat{\bar\gamma}_{11},\\
\hat{\bar a}_4&=3\hat\xi-16\hat{\bar\gamma}_3-4\hat{\bar\gamma}_4-64\hat{\bar\gamma}_7-4\hat{\bar\gamma}_8-16\hat{\bar\gamma}_9,
\end{aligned}
\qquad
\begin{aligned}
\hat{\bar a}_5&=3\hat\xi+8\hat{\bar\gamma}_3+20\hat{\bar\gamma}_4+32\hat{\bar\gamma}_7+20\hat{\bar\gamma}_8+44\hat{\bar\gamma}_9,\\
\hat{\bar a}_6&=6\hat\xi+40\hat{\bar\gamma}_3+28\hat{\bar\gamma}_4+16\hat{\bar\gamma}_7+100\hat{\bar\gamma}_8+40\hat{\bar\gamma}_9,\\
\hat{\bar a}_7&=\hat\xi-8\hat{\bar\gamma}_1+8\hat{\bar\gamma}_2,\\
\hat{\bar b}_1&=4\hat{\bar\omega}_1-3\hat{\bar\omega}_4,\\
\hat{\bar b}_2&=\hat{\bar\omega}_3+4\hat{\bar\omega}_5+\hat{\bar\omega}_6,\\
\hat{\bar b}_3&=-\hat{\bar\omega}_3+2\hat{\bar\omega}_5+5\hat{\bar\omega}_6.
\end{aligned}
\label{eq:dimless}
\end{equation}

\subsection{Coefficients of $k(\phi)$ \eqref{eq:kpoly_2}}\label{kphicoeffgen}

To write the coefficients of Eq.~\eqref{eq:kpoly_2} in a compact form we proceed in the following way. Since every entry of $\bar M (f)$ is linear in $f$, its determinant
and adjugate can be expanded as
\begin{equation}
-\det \bar M(f)
=
4 M_P^2 (\hat{d}_1f+\hat{d}_2f^2+\hat{d}_3f^3+\hat{d}_4f^4),
\label{eq:dhat_general}
\end{equation}
and
\begin{equation}
\operatorname{adj}\bar M (f)
=
\hat{\mathcal C}_0
+\hat{\mathcal C}_1 f
+\hat{\mathcal C}_2 f^2
+\hat{\mathcal C}_3 f^3.
\label{eq:Chat_general}
\end{equation}
The coefficient matrices and the denominator coefficients are therefore
\begin{equation}
\hat{\mathcal C}_n
=
\frac{1}{n!}
\left.
\frac{\partial^n}{\partial f^n}
\operatorname{adj}\bar M(f)
\right|_{f=0},
\qquad n=0,1,2,3,
\end{equation}
and
\begin{equation}
\hat{d}_n
=
-\frac{1}{4 \Mp^2 n!}
\left.
\frac{\partial^n}{\partial f^n}
\det\bar M(f)
\right|_{f=0},
\qquad n=1,\ldots,4.
\end{equation}
Using $\bar S=\Mp^2 f'(\phi)\,\hat{\bar s}$, with $\hat{\bar s}$ given in Eq.~\eqref{eq:split}, the numerator
coefficients are

\begin{equation}
\begin{aligned}
\hat{\kappa}_1&=-\frac{1}{2}\,
\hat{\bar s}^{\rm T}\hat{\mathcal C}_0 \hat{\bar s},
\qquad
\hat{\kappa}_2=-\frac{1}{2}\,
\hat{\bar s}^{\rm T}\hat{\mathcal C}_1 \hat{\bar s},
\\[4pt]
\hat{\kappa}_3&=-\frac{1}{2}\,
\hat{\bar s}^{\rm T}\hat{\mathcal C}_2 \hat{\bar s},
\qquad
\hat{\kappa}_4=-\frac{1}{2}\,
\hat{\bar s}^{\rm T}\hat{\mathcal C}_3 \hat{\bar s}.
\end{aligned}
\label{eq:kappahat_general}
\end{equation}
Thus
\begin{equation}
k(\phi)
=
\Mp^2f'(\phi)^2\,
\frac{
\hat{\kappa}_1+\hat{\kappa}_2 f+\hat{\kappa}_3f^2+\hat{\kappa}_4f^3
}{
4 f\big(\hat{d}_1+\hat{d}_2f+\hat{d}_3f^2+\hat{d}_4f^3\big)
},
\end{equation}
which is Eq.~\eqref{eq:kpoly_2}.

\section{Comparison with Ref. \cite{Andrei:2026xad}}
\label{app:coeff_3}

\subsection{Tensors}
 The conventions used here differ slightly from those of
Ref.~\cite{Andrei:2026xad}.
In particular, the ordering of the two lower indices of the affine
connection and of the distortion is reversed:
\begin{align}
\Gamma^{\rho}{}_{\mu\nu}
&=
\left.\Gamma^{\rho}{}_{\nu\mu}\right|_{\text{\cite{Andrei:2026xad}}},
&
N^{\rho}{}_{\mu\nu}
&=
\left.N^{\rho}{}_{\nu\mu}\right|_{\text{\cite{Andrei:2026xad}}}.
\end{align}
The torsion tensor itself has the same sign and index ordering, while the
non-metricity tensor differs by an overall sign. Therefore the relations for torsion and nonmetricity tensor and vectors are
\begin{equation}
\begin{array}{c@{\qquad\qquad}c}
\displaystyle
T^{\rho}{}_{\mu\nu}
=
\left. -2 S_{\mu\nu}{}^{\rho}\right|_{\text{\cite{Andrei:2026xad}}}
&
\displaystyle
Q_{\rho\mu\nu}
=
-\left.Q_{\rho\mu\nu}\right|_{\text{\cite{Andrei:2026xad}}}
\\[8pt]
\displaystyle
T_\mu
=
\left.2 S_\mu\right|_{\text{\cite{Andrei:2026xad}}}
&
\displaystyle
Q_\mu
=
-\left.Q_\mu\right|_{\text{\cite{Andrei:2026xad}}}
\\[8pt]
\displaystyle
\bar T_\mu
=
\left. - 2 t_\mu\right|_{\text{\cite{Andrei:2026xad}}}
&
\displaystyle
\tilde Q_\mu
=
-\left.q_\mu\right|_{\text{\cite{Andrei:2026xad}}}
\end{array}
\end{equation}

\subsection{Coefficients of the total lagrangian (8.1)--(8.3)}
The coefficients of the general metric-affine action in Eqs.~(8.1)--(8.3), on the left of the $=$, are related to those of Ref.~\cite{Andrei:2026xad}, on the right, by the following dictionary, with $\mathcal B=1$ in Ref.~\cite{Andrei:2026xad}:

\begin{equation}
\begin{array}{c@{\qquad\qquad}c@{\qquad\qquad}c@{\qquad\qquad}c@{\qquad\qquad}c}

\hat F=\dfrac{\mathcal A}{\kappa}
&
\hat V=\mathcal V
&
\hat\gamma_1=\dfrac{b_1}{8\kappa}
&
\hat\gamma_2=-\dfrac{b_2}{8\kappa}
&
\hat\gamma_3=\dfrac{a_1}{2\kappa}
\\[6pt]

\hat\gamma_4=\dfrac{a_2}{2\kappa}
&
\hat\gamma_5=\dfrac{c_1}{4\kappa}
&
\hat\gamma_6=\dfrac{b_3}{8\kappa}
&
\hat\gamma_7=\dfrac{a_3}{2\kappa}
&
\hat\gamma_8=\dfrac{a_4}{2\kappa}
\\[6pt]

\hat\gamma_9=\dfrac{a_5}{2\kappa}
&
\hat\gamma_{10}=-\dfrac{c_2}{4\kappa}
&
\hat\gamma_{11}=-\dfrac{c_3}{4\kappa}
&
\hat\omega_1=\dfrac{b_5}{8\kappa}
&
\hat\omega_2=-\dfrac{a_6}{2\kappa}
\\[6pt]

\hat\omega_3=\dfrac{c_6}{4\kappa}
&
\hat\omega_4=-\dfrac{b_4}{8\kappa}
&
\hat\omega_5=\dfrac{c_4}{4\kappa}
&
\hat\omega_6=\dfrac{c_5}{4\kappa}
&
\hat\beta'_1=\dfrac{\mathcal C_3}{4}
\\[6pt]

\hat\beta'_2=-\dfrac{\mathcal C_1}{2}
&
\hat\beta'_3=-\dfrac{\mathcal C_2}{2}
&
\hat\beta'_4=-\dfrac{\mathcal C_4}{4}
&
&
\end{array}
\end{equation}

\bibliographystyle{JHEP}
\bibliography{ref}
\begin{acknowledgments}
    This work was supported by the Estonian Research Council via the grants TARISTU24-TK10, TARISTU24-TK3, and the CoE program TK202 ``Foundations of the Universe''.
\end{acknowledgments}
\end{document}